\documentclass[aps,prc,twocolumn,floatfix,nofootinbib,showpacs,superscriptaddress]{revtex4-1}

\usepackage{amsmath,amsfonts,amssymb,bm}
\usepackage{graphicx}
\usepackage{color}
\definecolor{purple}{rgb}{0.5,0,0.5}
\definecolor{blue}{rgb}{0.0,0,0.9}

\begin{document}

\title{Commentary on rainbow-ladder truncation for excited states and exotics}

\author{Si-xue Qin}
\affiliation{Department of Physics, Center for High Energy Physics and State Key Laboratory of Nuclear Physics and Technology, Peking University, Beijing 100871, China}

\author{Lei Chang}
\affiliation{Physics Division, Argonne National Laboratory, Argonne,
Illinois 60439, USA}

\author{Yu-xin Liu}
\affiliation{Department of Physics, Center for High Energy Physics and State Key Laboratory of Nuclear Physics and Technology, Peking University, Beijing 100871, China}
\affiliation{Center of Theoretical Nuclear Physics, National
Laboratory of Heavy Ion Accelerator, Lanzhou 730000, China}

\author{Craig D.~Roberts}
\affiliation{Department of Physics, Center for High Energy Physics and State Key Laboratory of Nuclear Physics and Technology, Peking University, Beijing 100871, China}
\affiliation{Physics Division, Argonne National Laboratory, Argonne, Illinois 60439, USA}
\affiliation{Department of Physics, Illinois Institute of Technology, Chicago, Illinois 60616-3793, USA}

\author{David J.~Wilson}
\affiliation{Physics Division, Argonne National Laboratory, Argonne, Illinois 60439, USA}


\begin{abstract}
Ground-state, radially-excited and exotic scalar-, vector- and flavoured-pseudoscalar-mesons are studied in rainbow-ladder truncation using an interaction kernel that is consonant with modern DSE- and lattice-QCD results.  The inability of this truncation to provide realistic predictions for the masses of excited- and exotic-states is confirmed and explained.  On the other hand, its application does provide information that is potentially useful in proceeding beyond this leading-order truncation, e.g.: assisting with development of projection techniques that ease the computation of excited state properties; placing qualitative constraints on the long-range behaviour of the interaction kernel; and highlighting and illustrating some features of hadron observables that do not depend on details of the dynamics.
\end{abstract}


\maketitle

\section{Introduction}
Meson spectroscopy is a keystone of extant and forthcoming programmes at numerous facilities worldwide, e.g.: the Beijing Spectrometer; the COMPASS detector at CERN; Hall-D at Jefferson Laboratory; the Japan proton accelerator research complex (J-PARC); and the PANDA detector at GSI.  Each identifies an essentially identical primary motivation; namely, seeking answers  to two fundamental questions within the Standard Model: What matter is possible; and How is it constituted?  The subtext is quantum chromodynamics (QCD), the strongly-interacting part of the Standard Model, and the unique nature of the forces it seems to produce.  With QCD, Nature has prepared the sole known example of a strongly-interacting quantum field theory that is defined by degrees-of-freedom which cannot directly be detected; i.e., they are \emph{confined}.  One of the greatest challenges in modern physics is to comprehend and explain the phenomenon of confinement.

Following Ref.\,\cite{Wilson:1974sk}, confinement in mesons has typically been associated with a linearly rising potential between the quark-antiquark pair \cite{Eichten:1978tg}. There are sound reasons for using such potential model phenomenology in the study of heavy quarkonia \cite{Brambilla:2010cs}.  However, that is not true for light-quark systems.  The static potential measured in simulations of lattice-QCD is not related in any known way to the question of light-quark confinement.  Light-quark creation and annihilation effects are fundamentally nonperturbative.  Hence it is impossible in principle to compute a potential between two light quarks \cite{Bali:2005fu,Chang:2009ae}.  On the other hand, confinement can be related to the analytic properties of QCD's Schwinger functions \cite{Krein:1990sf,Roberts:1994dr,Roberts:2000aa,Maris:2003vk,Pennington:2005be,Roberts:2007ji,Chang:2011vu,%
Roberts:2011rr,Rodriguez-Quintero:2011}, so the question of light-quark confinement may be translated into the challenge of charting the infrared behavior of QCD's $\beta$-function.

To a large degree, this is also true of explaining dynamical chiral symmetry breaking (DCSB), a phenomenon which has an enormous impact on the measurable properties of mesons and baryons \cite{Chang:2011vu,Roberts:2011rr}.  It is known that DCSB; namely, the generation of mass \emph{from nothing}, does occur in QCD \cite{Bhagwat:2003vw,Bowman:2005vx,Bhagwat:2006tu}.  It arises primarily because a dense cloud of gluons comes to clothe a low-momentum quark \cite{Roberts:2007ji,Bhagwat:2007vx}.  This is readily seen by solving the Dyson-Schwinger equation (DSE) for the dressed-quark propagator; i.e., the gap equation.  However, the origin of the interaction strength at infrared momenta, which guarantees DCSB through the gap equation, is currently unknown.  This relationship ties confinement to DCSB.
The crucial role of DCSB means that reliable information about the $\beta$-function can only be obtained via a symmetry-preserving treatment of the bound-state problem that is capable of veraciously expressing DCSB.  The DSEs provide such a framework \cite{Roberts:1994dr,Roberts:2000aa,Maris:2003vk,Pennington:2005be,%
Roberts:2007ji,Chang:2011vu,Roberts:2011rr} and will be employed herein.

A considerable body of recent work (e.g., Refs.\,\cite{Roberts:2007ji,Maris:1997tm,Holl:2004fr,Holl:2005vu,Lucha:2006rq,McNeile:2006qy,Bhagwat:2007rj,%
Chang:2009zb,Fischer:2009jm,Krassnigg:2009zh,Chang:2011ei,Blank:2011qk,Chang:2011vu,Qin:2011dd}) has shown that in order to gain sensitivity to the long-range part of the interaction, one should minimally study the properties of mesons with significant rest-frame quark orbital angular momentum, such as scalar- and pseudovector-mesons, the radial excitations of pseudoscalar- and vector-mesons, and tensor mesons.   A challenging aspect of this problem is that the leading-order (rainbow-ladder) in the most widely used symmetry-preserving DSE truncation scheme \cite{Munczek:1994zz,Bender:1996bb} fails to adequately express the full power of DCSB in the kernels of the bound-state Bethe-Salpeter equations (BSEs) \cite{Chang:2009zb,Chang:2011ei,Chang:2010hb}.  Consequently, the results produced for systems other than ground-state flavoured-pseudoscalar- and vector-mesons have most often been qualitatively and quantitatively incorrect.

Is there any reason then to revisit the problem of the spectrum of excited and exotic mesons using the rainbow-ladder truncation?  The answer is ``no,'' if the goal is to extract quantitatively reliable information about the infrared behaviour of QCD's $\beta$-function.  On the other hand, the answer is ``yes,'' if one can exploit the truncation's simplicity in order to: identify features of excited and exotic states that are plausibly independent of the truncation; or techniques that can be useful in connection with more sophisticated truncations.  Such is our aim herein.

In Sec.\,\ref{gapBSE} we present the gap- and Bethe-Salpeter-equations in the symmetry-preserving rainbow-ladder truncation, explain the structure of their solutions and define their kernels.  Section~\ref{numerical} reports and interprets our numerical results, which include: masses and decay constants; an investigation of the relative importance of various Dirac structures within meson Bethe-Salpeter amplitudes; and an exploration of the pointwise behaviour and sign of the leading invariant amplitudes.  Section~\ref{epilogue} is an epilogue.

\section{Gap and Bethe-Salpeter equations}
\label{gapBSE}
The renormalised rainbow-gap- and ladder-Bethe-Salpeter-equations are, respectively:
\begin{eqnarray}
\nonumber S(p)^{-1} &=& Z_2 \,(i\gamma\cdot p + m^{\rm bm})\\
& +& Z_2^2 \int^\Lambda_\ell\!\! {\cal G}(\ell)
\ell^2 D_{\mu\nu}^{\rm free}(\ell)
\frac{\lambda^a}{2}\gamma_\mu S(p-\ell) \frac{\lambda^a}{2}\gamma_\nu ,
\label{rainbowdse} \rule{1em}{0ex}\\
\nonumber
\Gamma_M(k;P) &=&  - Z_2^2\int_q^\Lambda\!\!
{\cal G}((k-q)^2)\, (k-q)^2 \, D_{\mu\nu}^{\rm free}(k-q)\\
&& \times
\frac{\lambda^a}{2}\gamma_\mu S(q_+)\Gamma_M(q;P) S(q_-) \frac{\lambda^a}{2}\gamma_\nu ,
\label{ladderBSE}
\end{eqnarray}
where: we use a Euclidean metric \cite{Chang:2011vu}; $\int_\ell^\Lambda:=\int^\Lambda \!\! \mbox{\footnotesize $\frac{d^4 \ell}{(2\pi)^4}$}$ represents a Poincar\'e-invariant regularization of the integral, with $\Lambda$ the ultraviolet regularization mass-scale; $Z_2(\zeta,\Lambda)$ is the quark wavefunction renormalisation constant, whose location and strength in these equations may be understood from Refs.\,\cite{Bender:1996bb,Bloch:2002eq}; $D^{\rm free}_{\mu\nu}(\ell)$ is the Landau-gauge free-gauge-boson propagator;\footnote{Landau gauge is used for many reasons \protect\cite{Bashir:2009fv,Bashir:2011vg}, for example, it is: a fixed point of the renormalisation group; that gauge for which sensitivity to model-dependent differences between \emph{Ans\"atze} for the fermion--gauge-boson vertex are least noticeable; and a covariant gauge, which is readily implemented in numerical simulations of lattice regularised QCD \protect\cite{Cucchieri:2011aa}.}
one can choose $q_\pm=q\pm P/2$ without loss of generality in this Poincar\'e covariant approach; and
\begin{equation}
\label{GIR}
\ell^2 {\cal G}(\ell^2)
= \ell^2 {\cal G}_{\rm IR}(\ell^2) + 4\pi \tilde\alpha_{\rm pQCD}(\ell^2)
\end{equation}
specifies the interaction, with $\tilde\alpha_{\rm pQCD}(k^2)$ a bounded, monotonically-decreasing regular continuation of the perturbative-QCD running coupling to all values of spacelike-$\ell^2$, and ${\cal G}_{\rm IR}(\ell^2)$ an \emph{Ansatz} for the interaction at infrared momenta, such that ${\cal G}_{\rm IR}(\ell^2)\ll \tilde\alpha_{\rm pQCD}(\ell^2)$ $\forall \ell^2\gtrsim 2\,$GeV$^2$.  The form of ${\cal G}_{\rm IR}(\ell^2)$ determines whether confinement and/or DCSB are realised in solutions of the gap equation.

The solution of the gap equation is a dressed-quark propagator
\begin{equation}
 S(p) =\frac{1}{i \gamma\cdot p \, A(p^2,\zeta^2) + B(p^2,\zeta^2)}
= \frac{Z(p^2,\zeta^2)}{i\gamma\cdot p + M(p^2)}\,,
\label{SgeneralN}
\end{equation}
%
which is obtained from Eq.\,(\ref{rainbowdse}) augmented by a renormalisation condition.  A mass-independent scheme is a useful choice and can be implemented by fixing all renormalisation constants in the chiral limit.  Notably, the mass function, $M(p^2)=B(p^2,\zeta^2)/A(p^2,\zeta^2)$, is independent of the renormalisation point, $\zeta$; and the renormalised current-quark mass is given by
\begin{equation}
\label{mzeta}
m^\zeta = Z_m(\zeta,\Lambda) \, m^{\rm bm}(\Lambda) = Z_4^{-1} Z_2\, m^{\rm bm},
\end{equation}
wherein $Z_4$ is the renormalisation constant associated with the Lagrangian's mass-term. Like the running coupling constant, this ``running mass'' is a familiar concept.  However, it is not commonly appreciated that $m^\zeta$ is simply the dressed-quark mass function evaluated at one particular deep spacelike point; viz,
\begin{equation}
m^\zeta = M(\zeta^2)\,.
\end{equation}

The renormalisation-group invariant current-quark mass may be inferred via
\begin{equation}
\hat m_f = \lim_{p^2\to\infty} \left[\frac{1}{2}\ln \frac{p^2}{\Lambda^2_{\rm QCD}}\right]^{\gamma_m} M_f(p^2)\,,
\end{equation}
where $f$ specifies the quark's flavour, $\gamma_m = 12/(33-2 N_{f_\alpha})$: $N_{f_\alpha}$ is the number of quark flavours employed in computing the running coupling; and $\Lambda_{\rm QCD}$ is QCD's dynamically-generated renormalisation-group-invariant mass-scale.  The chiral limit is expressed by
\begin{equation}
\hat m_f = 0\,.
\end{equation}
Moreover,
\begin{equation}
\forall \zeta^2 \gg \Lambda_{\rm QCD}^2 , \;
\frac{M_{f_1}(p^2=\zeta^2)}{M_{f_2}(p^2=\zeta^2)}
=\frac{m_{f_1}^\zeta}{m^\zeta_{f_2}}=\frac{\hat m_{f_1}}{\hat m_{f_2}} .
\end{equation}

We would like to emphasise, however, that in the presence of DCSB the ratio $M_{f_1}(p^2)/M_{f_2}(p^2)$ is not independent of $p^2$: in the infrared; i.e., $\forall p^2 \lesssim \Lambda_{\rm QCD}^2$, it then expresses a ratio of constituent-like quark masses, which, for light quarks, are two orders-of-magnitude larger than their current-masses and nonlinearly related to them \cite{Holl:2005st,Flambaum:2005kc}.  (See, e.g., the discussion following Eq.\,(\ref{currentmasses}).)

The BSE is an eigenvalue problem for the meson masses-squared; i.e., in a given channel Eq.\,(\ref{ladderBSE}) has solutions only at particular, isolated values of $P^2=-m_M^2$.  At these values, solving the equation produces the associated meson's Bethe-Salpeter amplitude, which can then be used in the computation of observable properties.  Herein we consider\footnote{Masses and other properties of charge-neutral pseudoscalar mesons are affected by the non-Abelian anomaly.  In the BSE context, this is discussed in Ref.\,\protect\cite{Bhagwat:2007ha}.  Since the non-Abelian anomaly is a correction to rainbow-ladder truncation that is qualitatively different to the focus of our study, herein we specialise to flavoured pseudoscalars.}
flavoured-pseudoscalar-, scalar- and vector-meson ground-, radially-excited- and exotic-states, so that the following amplitudes arise:
\begin{eqnarray}
\label{GammaP}
\Gamma_{J^P=0^-}(k;P) &= &\sum_{i=1}^4 \gamma_5 \tau_{0^-}^i(k,P)\, F_{0^-}^i(k;P),\\
\label{GammaS}
\Gamma_{0^+}(k;P) &= &\sum_{i=1}^4 \tau_{0^+}^i(k,P)\, F_{0^+}^i(k;P),\\
\label{GammaV}
\Gamma_{1^-}(k;P) &= &\sum_{i=1}^8 \tau_{1^-}^i(k,P)\, F_{1^-}^i(k;P),
\end{eqnarray}
with ($a^T_\mu := a_\mu - P_\mu \, a\cdot P /P^2$)\\
\begin{subequations}
\begin{eqnarray}
\tau_{0^-}^1 &=& i \tau_{0^+}^1 = i \mbox{\boldmath $I$}_D,\;\\
\tau_{0^-}^2 &=& \gamma\cdot P,\;
\tau_{0^+}^2 = k\cdot P \,\tau_{0^-}^2 ,\\
%
\tau_{0^-}^3 &=& k \cdot P \,\tau_{0^+}^3,
\;\tau_{0^+}^3 = P^2 \gamma\cdot k - k\cdot P \gamma\cdot P,\\
%
%
\tau_{0^-}^4 &=& \tau_{0^+}^4 = \sigma_{\mu\nu} P_\mu k_\nu,\\
\tau_{1^-}^1 &=& i\gamma^T_\mu , \\
\tau_{1^-}^2 &=& i[3 k^T_\mu  \gamma\cdot k^T - \gamma^T_\mu k^T\cdot k^T], \\
\tau_{1^-}^3 &=& i k^T_\mu\,k\cdot P\,  \gamma\cdot P ,\\
\tau_{1^-}^4 &=&
i[\gamma^T_\mu \gamma\cdot P \,\gamma \cdot k^T + k^T_\mu \gamma\cdot P],\\
\tau_{1^-}^5&=& k^T_\mu,\\
\tau_{1^-}^6&=& k\cdot P [\gamma^T_\mu \gamma^T\cdot k -  \gamma\cdot k^T \gamma^T_\mu],\\
\nonumber \tau_{1^-}^7&=& (k^T)^2(\gamma^T_\mu \gamma\cdot P-\gamma\cdot P\gamma^T_\mu)\\
&&  - 2 k^T_\mu \gamma\cdot k^T \gamma\cdot P,\\
\tau_{1^-}^8 &=&  k^T_\mu \gamma\cdot k^T \gamma\cdot P.
\end{eqnarray}
\end{subequations}
The canonical normalisation condition (see, e.g., Eq.\,(27) in Ref.\,\cite{Maris:1997tm} or, more generally, Ref.\,\cite{LlewellynSmith:1969az}) constrains the bound-state to produce a pole with unit residue in the quark-antiquark scattering matrix.

It remains only to specify the interaction in order to proceed.  We use that explained in Ref.\,\cite{Qin:2011dd}; viz.,
\begin{equation}
\label{CalGQC}
{\cal G}(s) = \frac{8 \pi^2}{\omega^4} D \, {\rm e}^{-s/\omega^2}
+ \frac{8 \pi^2 \gamma_m \,  {\cal F}(s)}{\ln [ \tau + (1+s/\Lambda_{\rm QCD}^2)^2]},
\end{equation}
where: $\gamma_m = 12/25$, $\Lambda_{\rm QCD}=0.234\,$GeV; $\tau={\rm e}^2-1$; and ${\cal F}(s) = \{1 - \exp(-s/[4 m_t^2])\}/s$, $m_t=0.5\,$GeV.  This interaction preserves the one-loop renormalisation-group behavior of QCD in the gap- and Bethe-Salpeter equations \cite{Maris:1997tm}, and the infrared behaviour can serve to ensure confinement and DCSB.
Moreover, it is consistent with modern DSE and lattice studies, which indicate that the gluon propagator is a bounded, regular function of spacelike momenta that achieves its maximum value on this domain at $s=0$ \cite{Bowman:2004jm,Aguilar:2009nf,Aguilar:2010gm}, and the dressed-quark-gluon vertex does not possess any structure which can qualitatively alter this behaviour \cite{Skullerud:2003qu,Bhagwat:2004kj}.
Notably, as illustrated in Ref.\,\cite{Qin:2011dd}, the parameters $D$ and $\omega$ are not independent: with $D\omega=\,$constant, one can expect computed observables to be practically insensitive to $\omega$ on the domain $\omega\in[0.4,0.6]\,$GeV.

\section{Numerical results for bound-states properties}
\label{numerical}
\subsection{Ground states}
Using the method of Ref.\,\cite{Krassnigg:2009gd}, we solved the gap equation for light $u=d$ quarks and the $s$-quark, with their current-quark masses fixed by requiring that the pion and kaon BSEs produce $m_\pi \approx 0.138\,$GeV and $m_K\approx 0.496\,$GeV.  This is straightforward in rainbow-ladder truncation because there is no coupling between the separate gap equations and no feedback from the BSEs \cite{Cloet:2008fw}; and yields
\begin{equation}
\label{currentmasses}
m_{u=d}^\zeta = 3.4\,{\rm MeV}\,,\; m_s^\zeta=82\,{\rm MeV}
\end{equation}
quoted at our renormalisation point $\zeta=19\,$GeV, a value chosen to match the bulk of extant studies.
These values correspond to renormalisation-group-invariant masses of $\hat m_{u,d}=6\,$MeV, $\hat m_s=146\,$MeV, one-loop-evolved masses of $m_{u=d}^{1\,{\rm GeV}}=5\,$MeV, $m_{s}^{1\,{\rm GeV}}=129\,$MeV; and give $m_s/m_u=24$.  They are consequently comparable with contemporary estimates by other means \cite{Nakamura:2010zzi}.  NB.\ With $\omega=0.6\,$GeV, $M^E_s/M^E_u = 1.52 \ll \hat m_s/\hat m_u$, where the constituent-quark mass $M_f^E:=\{ s| s>0, s = M_f^2(s)\}$.

\begin{table}[t]
\begin{center}
\begin{tabular*}
{\hsize}
{l@{\extracolsep{0ptplus1fil}}
||l@{\extracolsep{0ptplus1fil}}
|l@{\extracolsep{0ptplus1fil}}
|l@{\extracolsep{0ptplus1fil}}
|l@{\extracolsep{0ptplus1fil}}}
%
%
%
$\omega$ & 0.4 & 0.5 & 0.6 & 0.7 \\\hline
$A(0)$ & 2.07 & 1.70 & 1.38 & 1.16 \\
$M(0)$ & 0.62 & 0.52 & 0.42 & 0.29 \\
$m_\pi$ & 0.139~ & 0.134~ & 0.136~ & 0.139~ \\
$f_\pi$ & 0.094 & 0.093 & 0.090 & 0.081\\
$\rho_\pi^{1/2}$ & 0.49 & 0.49 & 0.49 & 0.48\\
$m_K$ & 0.496 & 0.495 & 0.497 & 0.503 \\
$f_K$  & 0.11 & 0.11 & 0.11 & 0.10 \\
$\rho_K^{1/2}$ & 0.55 & 0.55 & 0.55 & 0.55\\
$m_\sigma$ & 0.67 & 0.65 & 0.59 & 0.46 \\
$\rho_\sigma^{1/2}$ & 0.53 & 0.53 & 0.51 & 0.48\\
$m_\kappa$ & 0.89 & 0.88 & 0.85 & 0.77 \\
%
$f_{\kappa^+}$  & 0.035 & 0.036 & 0.037 & 0.042~ \\
%
$\rho_\kappa^{1/2}$ & 0.59 & 0.59 & 0.58 & 0.56\\
$m_\rho$ & 0.76 & 0.74 & 0.72 & 0.67 \\
$f_\rho$ & 0.14 & 0.15 & 0.14 & 0.12 \\
$m_\phi$ & 1.09 & 1.08 & 1.07 & 1.05 \\
$f_\phi$ & 0.19 & 0.19 & 0.19 & 0.18 \\
\end{tabular*}
\end{center}
\caption{\label{tableresults}
Results obtained using the interaction in Eq.\,(\protect\ref{CalGQC}) with $D\omega = (0.8\,{\rm GeV})^3$.  The current-quark masses at $\zeta=19\,$GeV are given in Eq.\,(\ref{currentmasses}).  Dimensioned quantities are reported in GeV.  For comparison, some experimental values are \protect\cite{Nakamura:2010zzi}: $f_\pi =0.092\,$GeV, $m_\pi = 0.138\,$GeV; $f_K=0.113\,$GeV, $m_K=0.496\,$GeV; $f_\rho=0.153\,$GeV, $m_\rho=0.777\,$GeV; and $f_\phi=0.168\,$GeV, $m_\phi=1.02\,$GeV.
NB.\ The scalar mesons listed here are not directly comparable with the lightest scalars in the hadron spectrum because the rainbow-ladder truncation is \emph{a priori} known to be a poor approximation in this channel: nonresonant corrections \protect\cite{Chang:2009zb,Chang:2011ei} and resonant final-state interactions are both important \protect\cite{Holl:2005st}.}
\end{table}

In Table~\ref{tableresults} we report selected results related to ground-state pseudoscalar-, scalar- and vector-mesons.  The meson masses are obtained in solving the BSEs.  Regarding the other meson quantities, in terms of the canonically normalised Bethe-Salpeter amplitudes and with
\begin{equation}
\chi_{J_{12}^P}(k;P) = S_{f_1}(k_+) \Gamma_{J^P}(k;K) S_{f_2}(k_-) ,
\end{equation}
where $f_1$, $f_2$ are the meson's valence-quark and \mbox{-antiquark}, respectively, one has \cite{Maris:1997tm,Roberts:2011IHC,Ivanov:1998ms}
\begin{eqnarray}
f_{0_{12}^-} P_\mu
& = & Z_2\; {\rm tr}_{\rm CD}
\int_k^\Lambda i\gamma_5\gamma_\mu \chi_{0_{12}^-}(k;P) \,, \label{fpigen}\\
i\rho_{0_{12}^-}^\zeta
& = & Z_4\; {\rm tr}_{\rm CD}
\int_k^\Lambda \gamma_5 \chi_{0_{12}^-}(k;P)  \,,\label{rhogen}\\
f_{0_{12}^+} P_\mu
& = & Z_2\; {\rm tr}_{\rm CD}
\int_k^\Lambda i\gamma_\mu \chi_{0_{12}^+}(k;P) \,, \label{fkappagen}\\
\rho^\zeta_{0_{12}^+}
& = & - Z_4\, {\rm tr}_{\rm CD}\!\!\!
\int_k^\Lambda \chi_{0_{12}^+}(k;P) \,, \label{rhoSgen}\\
f_{1_{12}^-} m_{1_{12}^-}
& = & \mbox{\small $\frac{1}{3}$} \, Z_2\; {\rm tr}_{\rm CD}
\int_k^\Lambda \gamma_\mu \chi_{1_{12}^-}(k;P) \,. \label{frhogen}
\end{eqnarray}
The Table confirms that, with $D\omega=\,$constant, observable properties of ground-state scalar-, vector- and flavoured-pseudoscalar-mesons computed with Eq.\,(\ref{CalGQC}) are practically insensitive to variations of $\omega \in [0.4,0.6]\,$GeV.

It is noteworthy, and readily verified using entries in the Table, that the pseudoscalar- and scalar- meson masses satisfy the following identities, exact in QCD \cite{Maris:1997tm,Roberts:2011IHC}:\footnote{Notwithstanding complexities associated with the structure of light-quark scalars \protect\cite{Holl:2005st,Pelaez:2006nj,RuizdeElvira:2010cs}, the identity written here applies to any scalar meson that can be produced via $e^+ e^-$ annihilation.  It is not of experimental significance, however, if the pole is deep in the complex plane.}
\begin{eqnarray}
\label{gmorP}
f_{0_{12}^-} m_{0_{12}^-}^2 &=& (m_{f_1}^\zeta + m_{f_2}^\zeta) \rho_{0_{12}^-}^\zeta, \\
f_{0_{12}^+} m_{0_{12}^+}^2 &=& -(m_{f_1}^\zeta - m_{f_2}^\zeta) \rho_{0_{12}^+}^\zeta.
\end{eqnarray}
Furthermore, the products $f_{0_{12}^\pm} \rho_{0_{12}^\pm}$ describe in-meson condensates \cite{Maris:1997tm,Roberts:2011IHC,Brodsky:2010xf}.

\subsection{Radial excitations and exotics}
In addition to properties of the ground-states, we have computed selected quantities associated with $J=0,1$ radial excitations and exotics.  In the Poincar\'e covariant DSE treatment, exotic states appear as poles in vertices generated by interpolating fields with ``unnatural time-parity'' \cite{Burden:2002ps}.  Results are presented in Table~\ref{tableradial}.  The last column in the Table was prepared as follows.  We fitted the entries in each row to both $m(\omega)=\,$constant and
\begin{equation}
\label{eq:momega}
m(\omega) = \omega (c_0 + c_1 \omega),
\end{equation}
then computed the standard-deviation of the relative error in each fit, $\sigma_0$ for the constant and $\sigma_2$ for Eq.\,(\ref{eq:momega}), and finally formed the ratio: $\sigma_{20}=\sigma_2/\sigma_0$.

\begin{table}[t]
\begin{center}
\begin{tabular*}
{\hsize}
{l@{\extracolsep{0ptplus1fil}}
||l@{\extracolsep{0ptplus1fil}}
|l@{\extracolsep{0ptplus1fil}}
|l@{\extracolsep{0ptplus1fil}}
|l@{\extracolsep{0ptplus1fil}}}
$\omega$ & 0.4 & 0.5 & 0.6 & $\sigma_{20}$\\\hline
$m_\pi$ & 0.214~ & 0.155~ & 0.147~ & 0.83\;\;\\ 
$m_{0^{--}}$ & 0.814 & 0.940 & 1.053 & 0.03\\ 
$m_{\pi_1} $ & 1.119 & 1.283 & 1.411 & 0.02\\ 
$m_\sigma$ & 0.970 & 0.923 & 0.913 & 1.25 \\ 
$m_{0^{+-}}$ & 1.186 & 1.252 & 1.323 & 0.34\\ 
$m_{\sigma_1}$ & 1.358 & 1.489 & 1.575 & 0.14 \\ 
$m_\rho$ & 1.088 & 1.046 & 1.029 & 1.22\\ 
$m_{1^{-+}}$ & 1.234 & 1.277 & 1.318 & 0.60\\ 
$m_{\rho_1}$ & 1.253 & 1.260 & 1.303 & 0.03\\
\end{tabular*}
\end{center}
\caption{\label{tableradial}
Masses obtained with Eq.\,(\protect\ref{CalGQC}), $D\omega = (1.1\,{\rm GeV})^3$.  The subscript ``1'' indicates first radial excitation.  The last column measures sensitivity to variations in $r_\omega:=1/\omega$:  $\sigma_{20}\ll 1$ indicates strong sensitivity; and $\sigma_{20} \approx 1$, immaterial sensitivity. 
Dimensioned quantities reported in GeV.
}
\end{table}

In preparing the table we used $D\omega = (1.1\,{\rm GeV})^3$.  This has the effect of inflating the $\pi$- and $\rho$-meson ground-state masses to a point wherefrom corrections to rainbow-ladder truncation can plausibly return them to the observed values \cite{Eichmann:2008ae,Roberts:2011cf}.  It is therefore notable that, in contrast to Table~\ref{tableresults}, the value reported for $m_\sigma$ in Table~\ref{tableradial} matches estimates for the mass of the dressed-quark-core component of the $\sigma$-meson obtained using unitarised chiral perturbation theory \cite{Pelaez:2006nj,RuizdeElvira:2010cs}.

A comparison between the $\omega$-dependence of ground-state properties and those of excited- and exotic-states was drawn in Ref.\,\cite{Qin:2011dd} and we only summarise it here.  Ground-state masses of light-quark pseudoscalar- and vector-mesons are quite insensitive to $\omega\in[0.4,0.6]\,$GeV.  Any minor variation is described by a decreasing function.  In the case of exotics and radial excitations, the variation with $\omega$ is described by an increasing function and the variation is usually significant.  This is readily understood.  The quantity $r_\omega:=1/\omega$ is a length-scale that measures the range over which the infrared part of Eq.\,(\ref{GIR}), ${\cal G}_{\rm IR}$, is active.  For $\omega=0$ this range is infinite, but it decreases with increasing $\omega$.  One expects exotic- and excited-states to be more sensitive to long-range features of the interaction than ground-states and, additionally, that their masses should increase if the magnitude and range of the strong piece of the interaction is reduced because there is less binding energy.

Table~\ref{tableradial} confirms a known fault with the rainbow-ladder truncation; viz., whilst it binds in exotic channels, it produces masses that are too light, just as it does for axial-vector mesons.
It is similarly noticeable that $m_{\pi_1}$ is far more sensitive to variations in $\omega$ than is $m_{\rho_1}$; and although $m_{\pi_1}<m_{\rho_1}$ for $\omega=0.4\,$GeV, the ordering is rapidly reversed.  Thus, in conflict with experiment, one usually finds $m_{\pi_1}>m_{\rho_1}$ in rainbow-ladder truncation.
This, too, is a property of the truncation, which is insensitive to the details of ${\cal G}(k^2)$; e.g., the same ordering is obtained with a momentum-independent interaction \cite{Roberts:2011cf}.

\begin{figure}[t]

\centerline{\includegraphics[clip,width=0.46\textwidth]{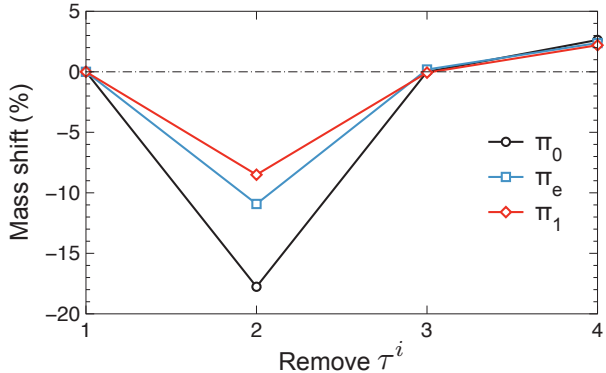}}

\caption{\label{FigPIA}
Pseudoscalar mesons.  Relative difference between the mass computed with all the amplitudes in Eq.\,(\protect\ref{GammaP}) and that obtained when the identified $i\geq 2$ amplitude is omitted: \emph{circles} -- ground-state pion; \emph{squares} -- $J^{PC}=0^{--}$ exotic; and \emph{diamonds} -- first pseudoscalar radial excitation.  In all cases, $\omega=0.6\,$GeV, $D\omega = (1.1\,{\rm GeV})^3$.  There is only minor quantitative variation with $\omega\in [0.4,0.6]\,$GeV.  NB.\ The $i=1$ amplitude is never omitted, it specifies the reference value.}
\end{figure}

\subsection{Structure of bound states}
In order to develop insight, both into the structure of excited- and exotic-states, and for progressing beyond rainbow-ladder truncation, it is useful to know which of the invariant amplitudes in Eqs.\,(\ref{GammaP})-(\ref{GammaV}) are dominant.  One useful measure of an amplitude's importance is the contribution it makes to a given meson's mass.  Figure~\ref{FigPIA} displays the result for pseudoscalar mesons: in all cases a good approximation is obtained by retaining $F_{0^-}^1$ and $F_{0^-}^2$.  This outcome is in agreement with extant ground-state computations \cite{Maris:1997tm} but extends those rainbow-ladder conclusions to excited- and exotic-states.  Evidently, there is little here to distinguish between the exotic and the radial excitation. Curiously, $F_{0^-}^2$ plays a role of similar magnitude in each state and the amplitudes $F_{0^-}^3$ and $F_{0^-}^4$ are always largely unimportant.  These last two, in this instance small, amplitudes are those most directly associated with nonzero quark orbital angular momentum in the meson's rest-frame.

\begin{figure}[t]

\centerline{\includegraphics[clip,width=0.46\textwidth]{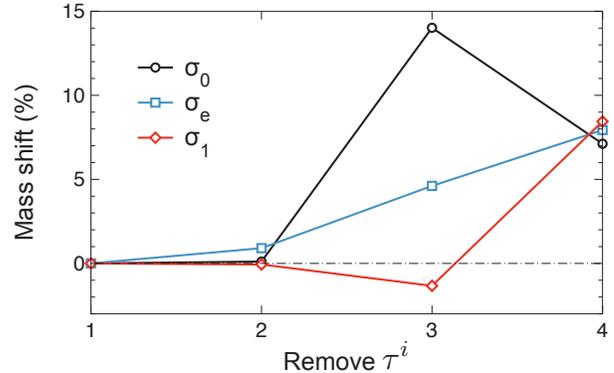}}

\caption{\label{FigSIA}
Scalar mesons.  Relative difference between the mass computed with all the amplitudes in Eq.\,(\protect\ref{GammaS}) and that obtained when the identified $i\geq 2$ amplitude is omitted: \emph{circles} -- ground-state $u=d$ scalar; \emph{squares} -- $J^{PC}=0^{+-}$ exotic; and \emph{diamonds} -- first pseudoscalar radial excitation.  In all cases, $\omega=0.6\,$GeV, $D\omega = (1.1\,{\rm GeV})^3$.  There is only minor quantitative variation with $\omega\in [0.4,0.6]\,$GeV.  NB.\ The $i=1$ amplitude is never omitted, it specifies the reference value.}
\end{figure}

For scalar mesons, on the other hand, one reads from Fig.\,\ref{FigSIA} that $F_{0^+}^1$, $F_{0^+}^3$ and $F_{0^+}^4$ should be included if a reliable approximation is to be obtained.  The latter two amplitudes are directly associated with significant rest-frame quark orbital angular momentum.  Notably, in quantum mechanical models, scalar mesons are identified as $^3 \!P_0$ states, in contrast to $^1\! S_0$ for pseudoscalar mesons.

\begin{figure}[t]

\centerline{\includegraphics[clip,width=0.46\textwidth]{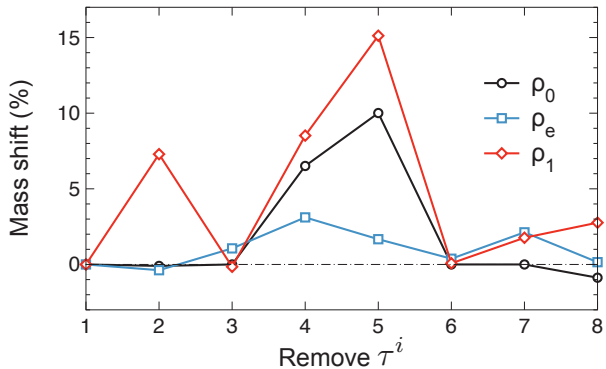}}

\caption{\label{FigVIA}
Vector mesons.  Relative difference between the mass computed with all the amplitudes in Eq.\,(\protect\ref{GammaV}) and that obtained when the identified $i\geq 2$ amplitude is omitted: \emph{circles} -- ground-state $u=d$ vector; \emph{squares} -- $J^{PC}=1^{-+}$ exotic; and \emph{diamonds} -- first vector radial excitation.  In all cases, $\omega=0.6\,$GeV, $D\omega = (1.1\,{\rm GeV})^3$.  Whilst there are quantitative changes with $\omega$, the pattern of amplitude importance is unchanged.  NB.\ The $i=1$ amplitude is never omitted, it specifies the reference value.}
\end{figure}

The vector meson ($^3\! S_1$) situation is displayed in Fig.\,\ref{FigVIA}.  In agreement with Ref.\,\cite{Maris:1999nt}, a good approximation for the vector-meson ground-state is obtained by retaining $F_{1^-}^1$, $F_{1^-}^4$, $F_{1^-}^5$.  The last two amplitudes are associated with $P$-wave components in the rest-frame.  However, for the first radial excitation, $F_{1^-}^2$ is also important: this amplitude is directly associated with a $D$-wave component in the radially-excited vector-meson's rest frame.
These observations suggest that a BSE might be built which projects selectively onto the first radially excited state.

The additional information contained in these figures indicates that the shortcomings identified above, of the rainbow-ladder truncation for states other than ground-state vector- and flavoured-pseudoscalar-mesons, can be attributed to this truncation's inadequate expression in the Bethe-Salpeter kernels of effects which in quantum mechanics would be described as spin-orbit interactions.   Namely, treating the quark-gluon vertex as effectively bare in both the gap- and Bethe-Salpeter-equations leads to omission of critically important helicity-flipping interactions that are dramatically enhanced by DCSB, as discussed in Refs.\,\cite{Chang:2009zb,Chang:2010hb,Chang:2011ei}.

One may readily expand on this.  For example, vector meson bound states possess nonzero magnetic- and quadrupole-moments \cite{Roberts:2011wy}.  This fact, Fig.\,\ref{FigVIA} and the associated discussion together indicate that there is appreciably more dressed-quark orbital angular momentum within these states than within pseudoscalar mesons.  Hence, spin-orbit repulsion could significantly boost $m_{\rho_1}$ and thereby produce the correct level ordering; viz., $m_{\rho_1}>m_{\pi_1}$.
Moreover, since exotic states appear as poles in vertices generated by interpolating fields with ``unnatural time-parity,'' the importance of orbital angular momentum within these states is magnified.
These comments apply with equal force to tensor mesons, which cannot be formed without rest-frame quark orbital angular momentum.

At present the best hope for a realistic description of the meson spectrum within a Poincar\'e covariant approach\footnote{A lattice-QCD perspective on the meson spectrum may be drawn from Ref.\,\protect\cite{Dudek:2011bn}.} is provided by the essentially nonperturbative DSE truncation scheme whose use is illustrated most fully in Ref.\,\cite{Chang:2011ei}.  That symmetry-preserving scheme deeply embeds effects associated with DCSB into the Bethe-Salpeter kernel.

\begin{figure}[t]
\hspace*{-1ex}\begin{minipage}[t]{0.5\textwidth}
\begin{minipage}[t]{0.48\textwidth}
\leftline{\includegraphics[width=1.01\textwidth]{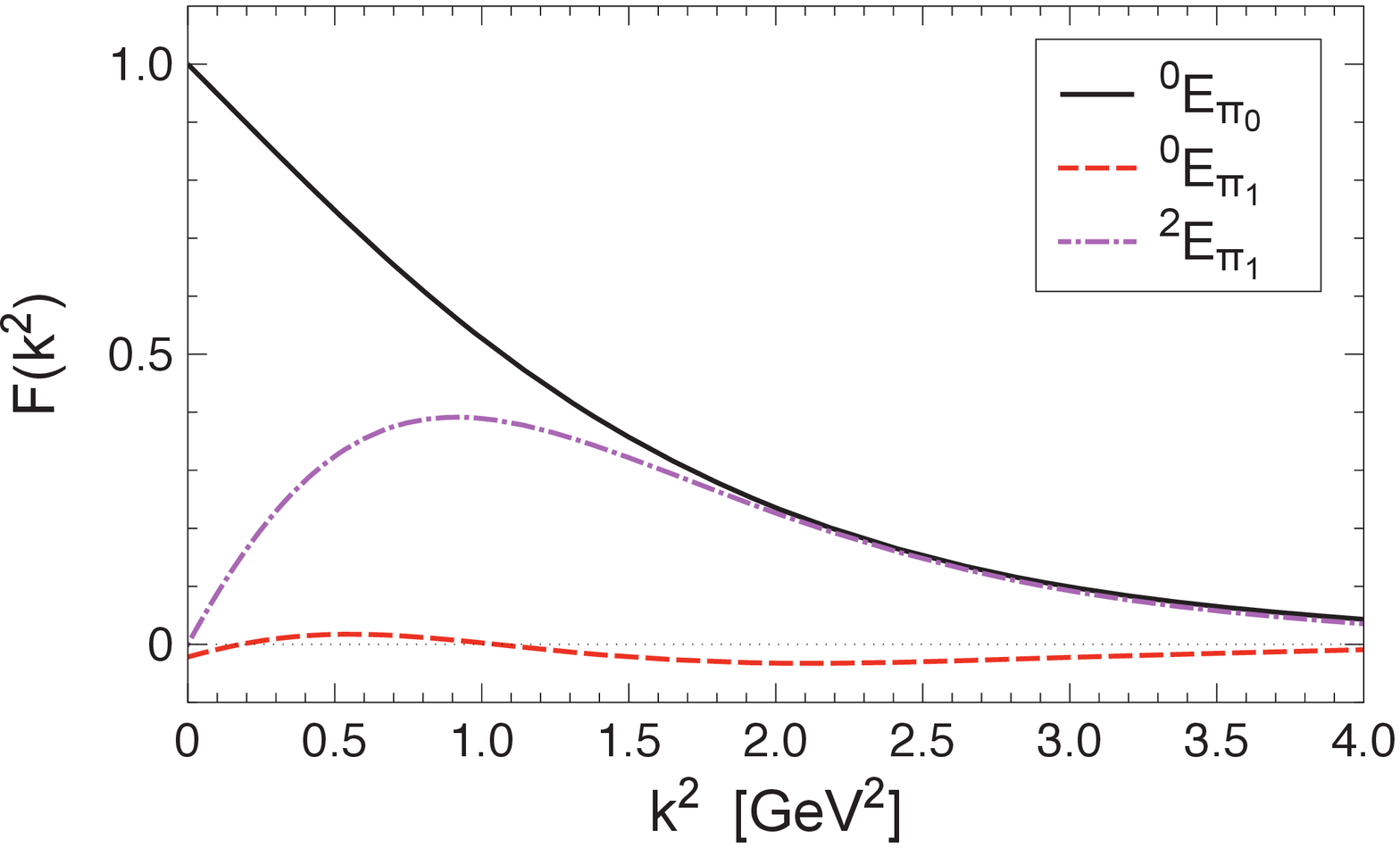}}
\end{minipage}
\begin{minipage}[t]{0.48\textwidth}
\rightline{\includegraphics[width=1.01\textwidth]{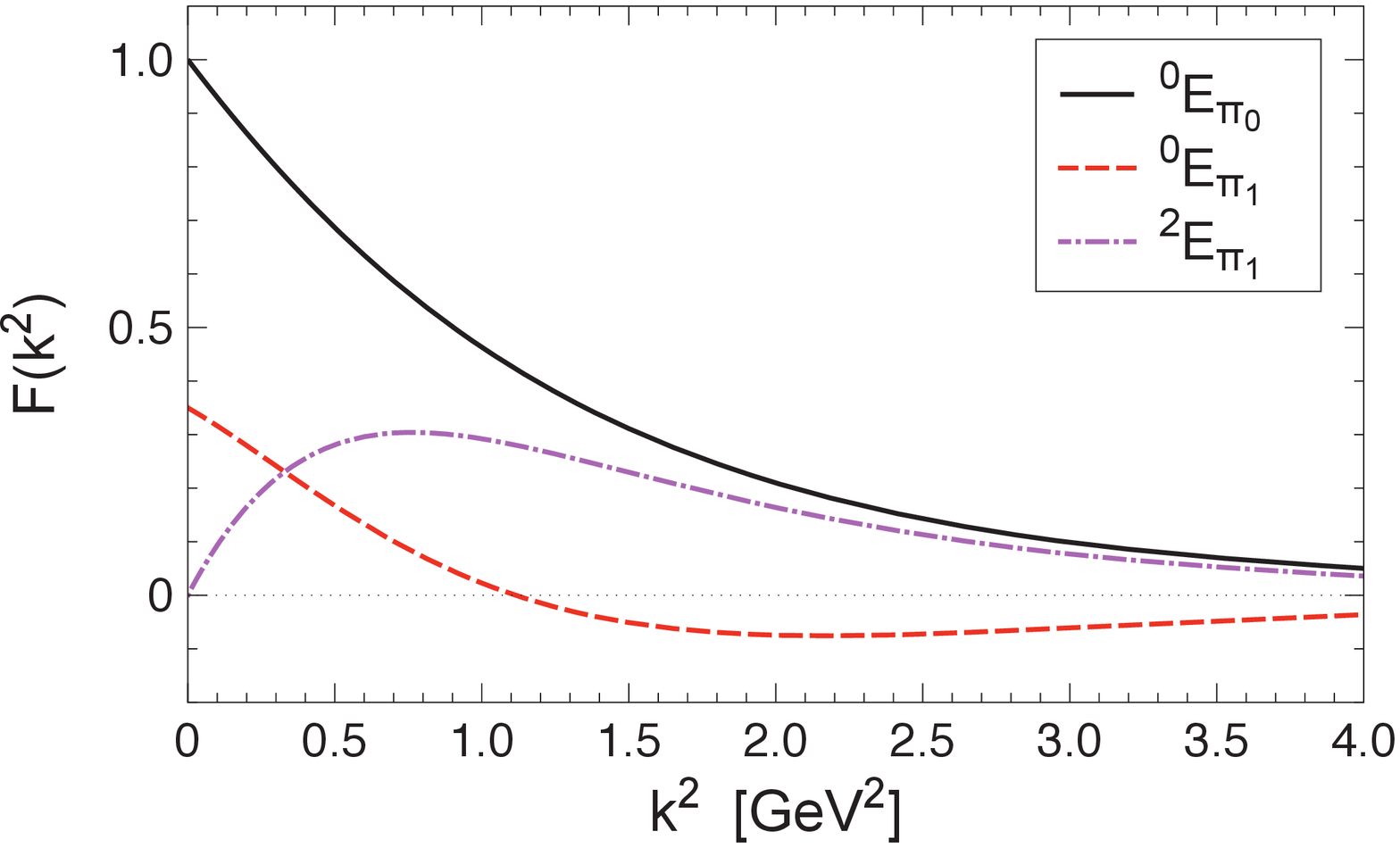}}
\end{minipage}\vspace*{3ex}

\begin{minipage}[t]{0.48\textwidth}
\leftline{\includegraphics[width=1.01\textwidth]{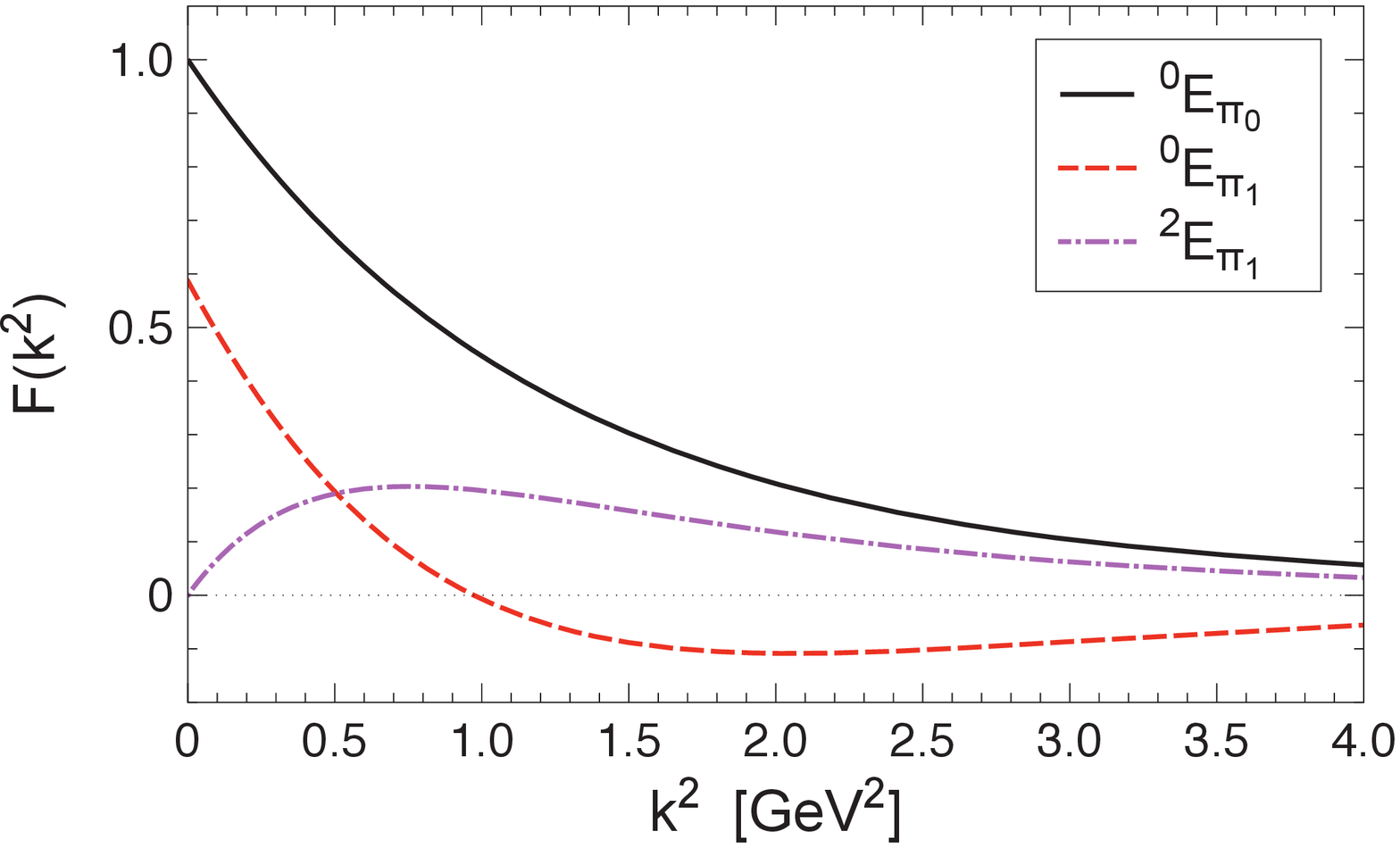}}
\end{minipage}
\begin{minipage}[t]{0.48\textwidth}
\rightline{\includegraphics[width=1.01\textwidth]{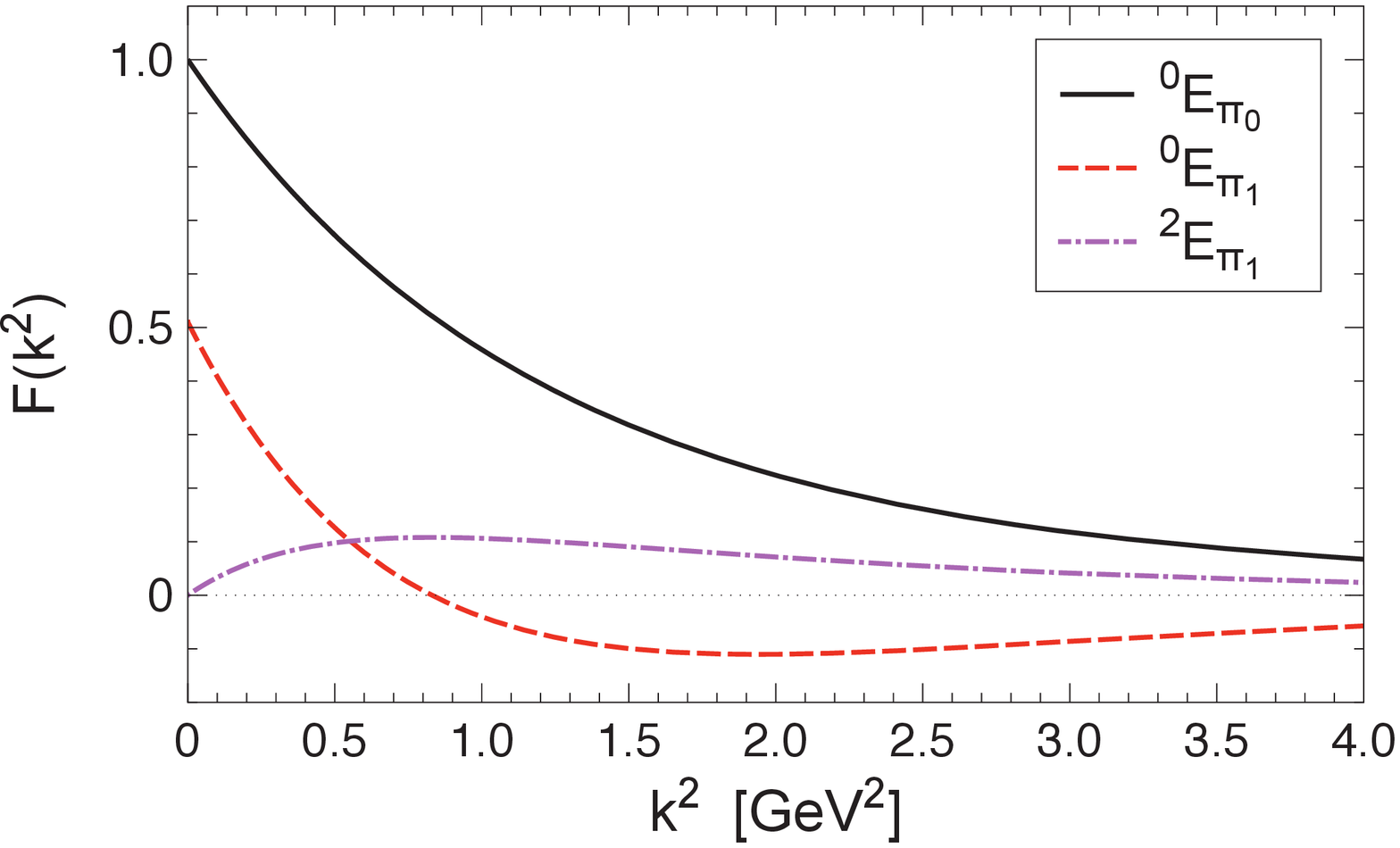}}
\end{minipage}\vspace*{3ex}

\begin{minipage}[t]{0.48\textwidth}
\leftline{\includegraphics[width=1.01\textwidth]{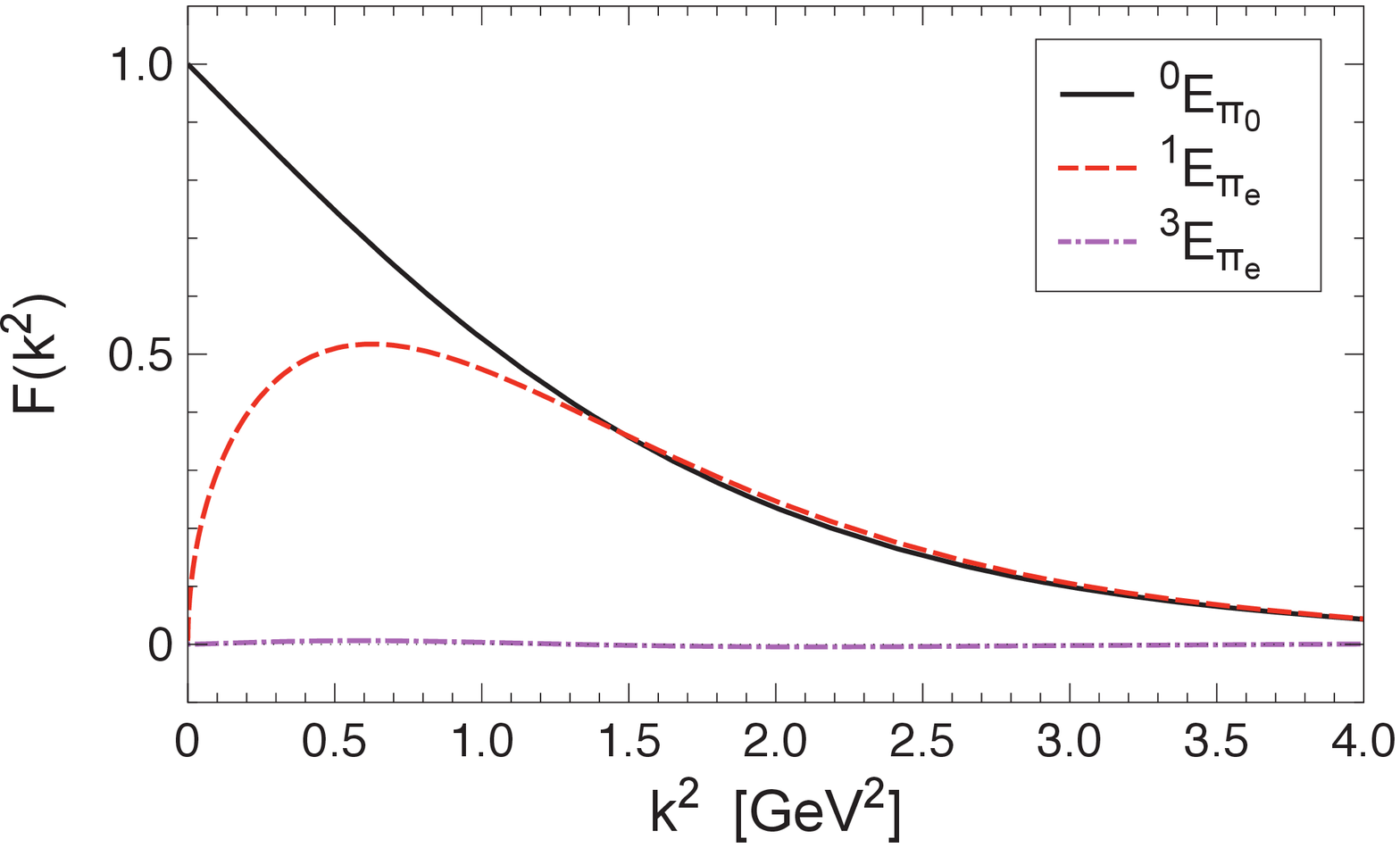}}
\end{minipage}
\begin{minipage}[t]{0.48\textwidth}
\rightline{\includegraphics[width=1.01\textwidth]{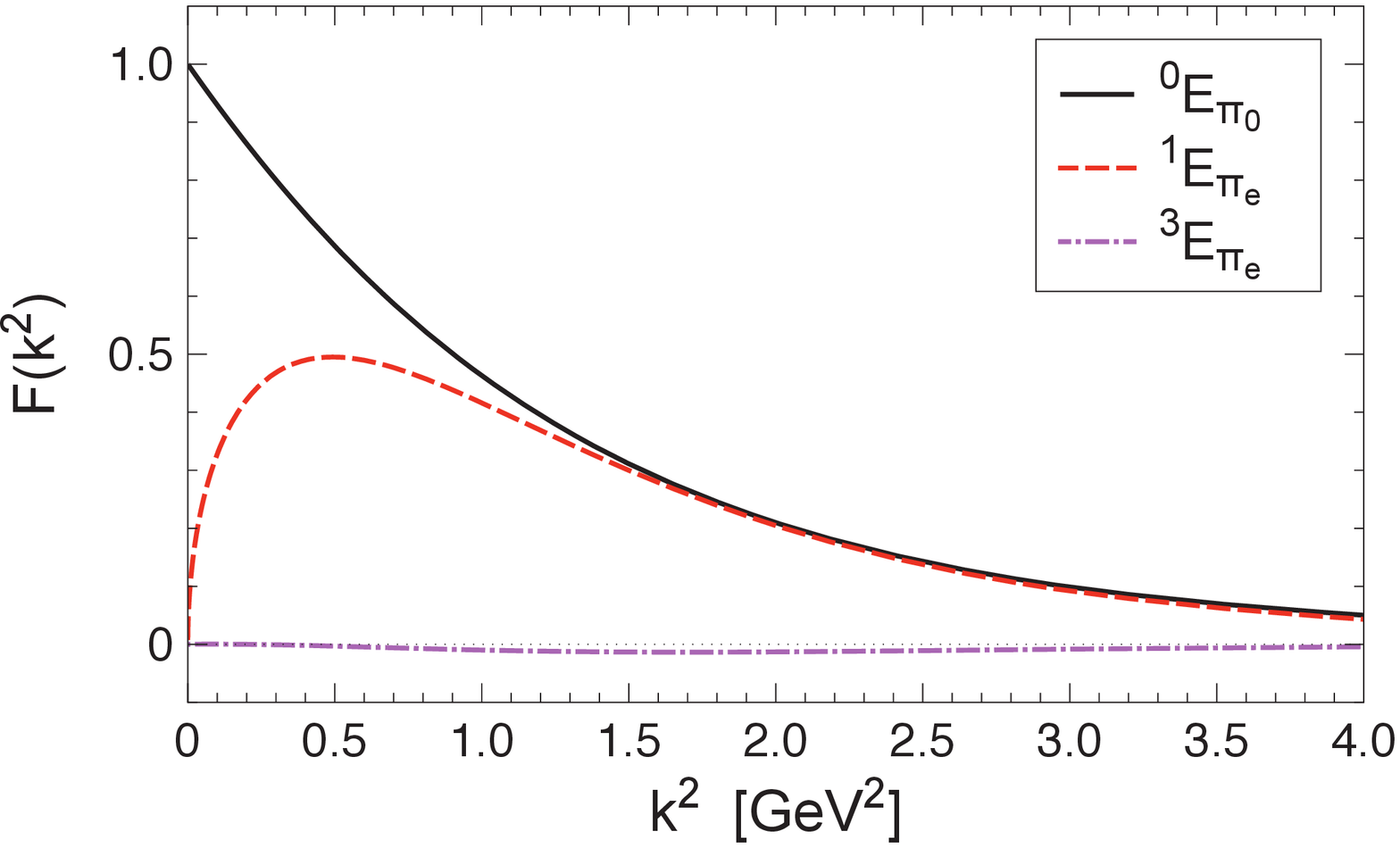}}
\end{minipage}\vspace*{3ex}

\begin{minipage}[t]{0.48\textwidth}
\leftline{\includegraphics[width=1.01\textwidth]{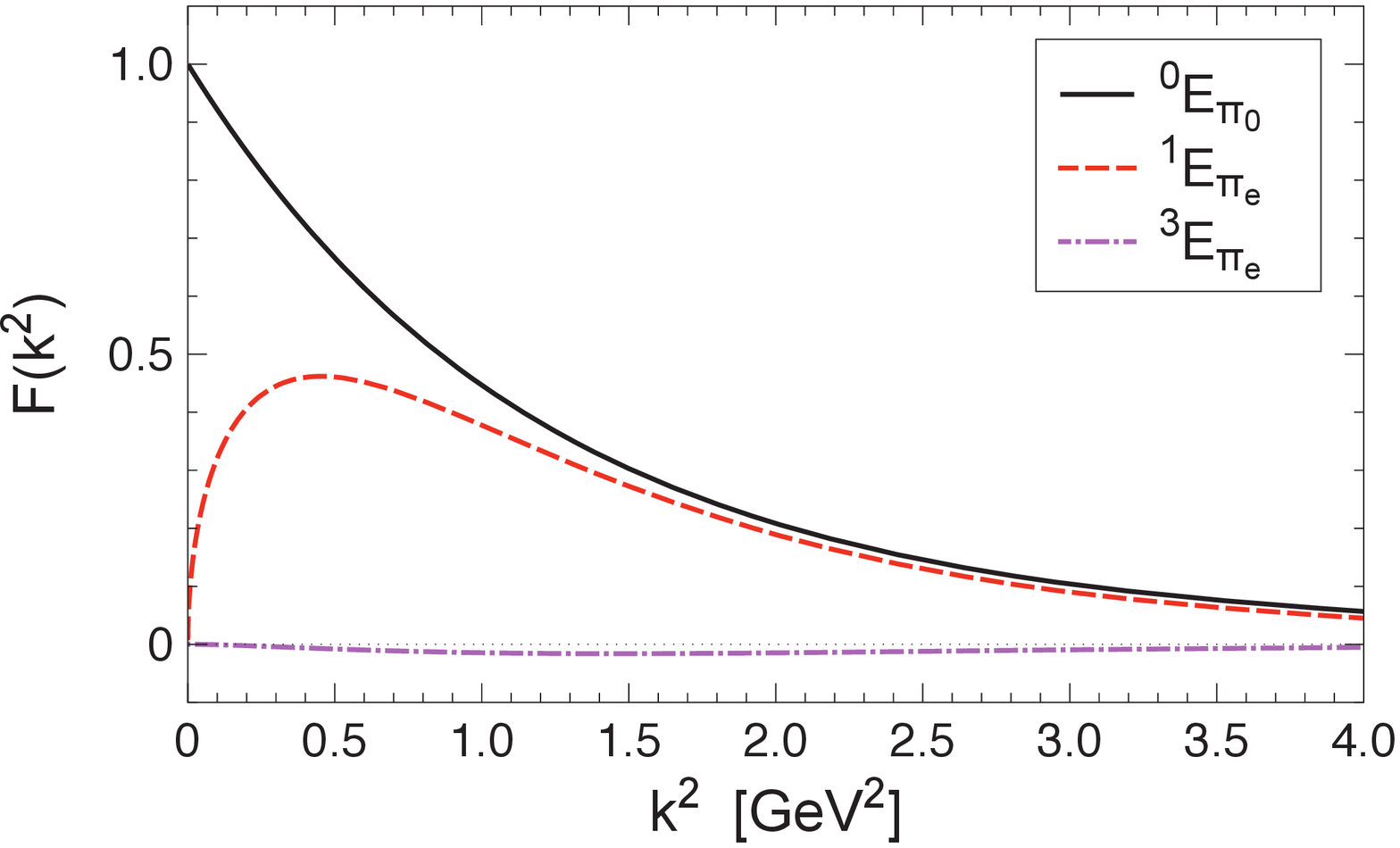}}
\end{minipage}
\begin{minipage}[t]{0.48\textwidth}
\rightline{\includegraphics[width=1.01\textwidth]{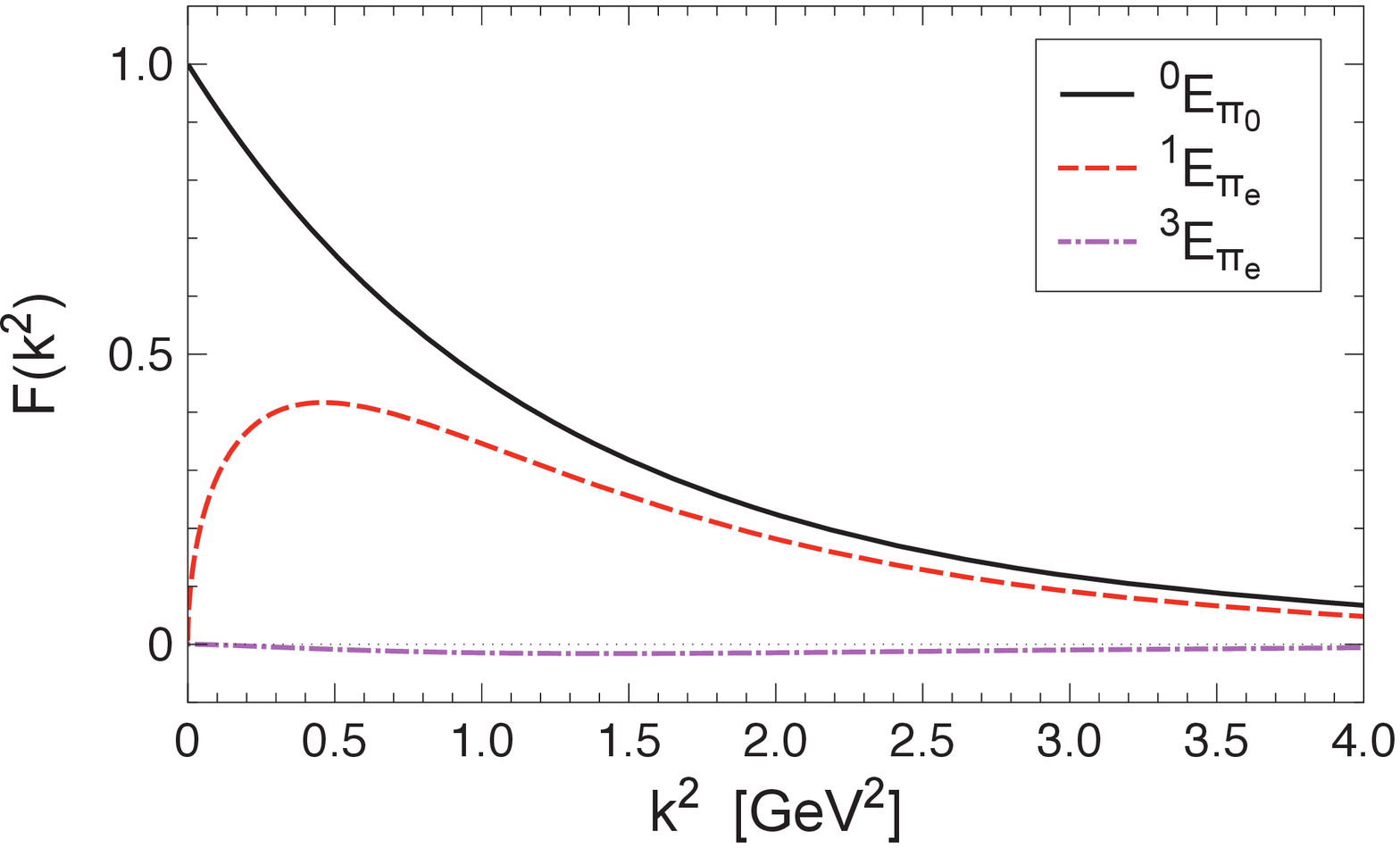}}
\end{minipage}
\end{minipage}

\caption{\label{pionBSA} Pseudoscalar mesons.  $\omega$-dependence of low-order Chebyshev-projections of leading invariant amplitude for ground-, radially-excited- and exotic-states: \emph{upper four panels}, ground and radial; \emph{lower four panels}, ground and exotic.  In all panels, \emph{solid} -- zeroth moment, ground-state; \emph{dashed} -- leading moment, comparison state; \emph{dash-dot} -- subleading moment, comparison state.  \emph{Row-1, left}, $\omega=0.4\,$GeV; \emph{Row-1, right}, $\omega=0.5\,$GeV; \emph{Row-2, left}, $\omega=0.6\,$GeV; and \emph{Row 2, right}, $\omega=0.7\,$GeV.  This pattern is repeated in the next two rows.  The normalisation is chosen such that $^0\!E_{\pi_0}(p^2=0)=1$; and $D\omega = (1.1\,{\rm GeV})^3$.}
\end{figure}

\subsection{Connecting amplitudes with observables}
\begin{figure}[t]
\hspace*{-1ex}\begin{minipage}[t]{0.5\textwidth}
\begin{minipage}[t]{0.48\textwidth}
\leftline{\includegraphics[width=1.01\textwidth]{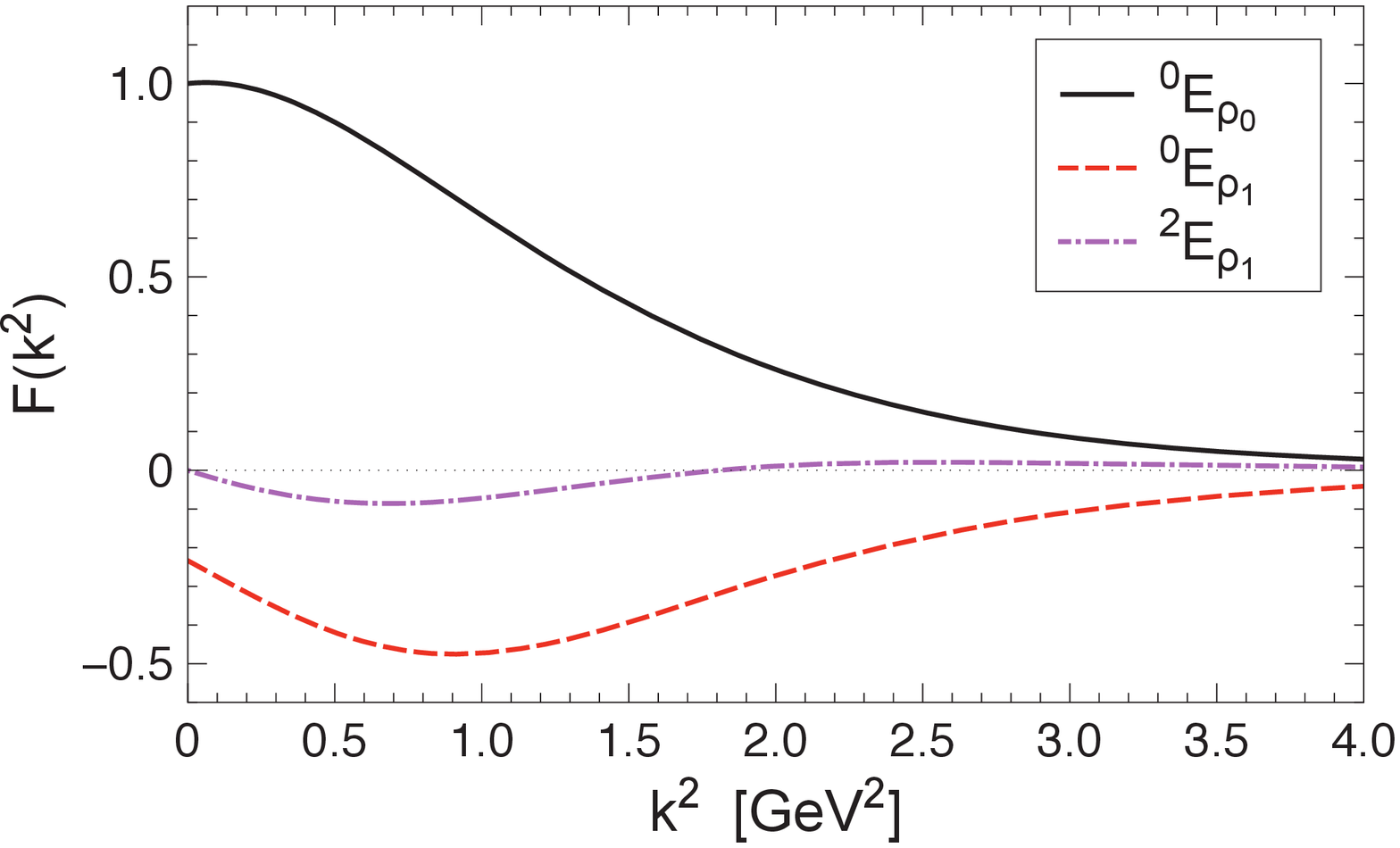}}
\end{minipage}
\begin{minipage}[t]{0.48\textwidth}
\rightline{\includegraphics[width=1.01\textwidth]{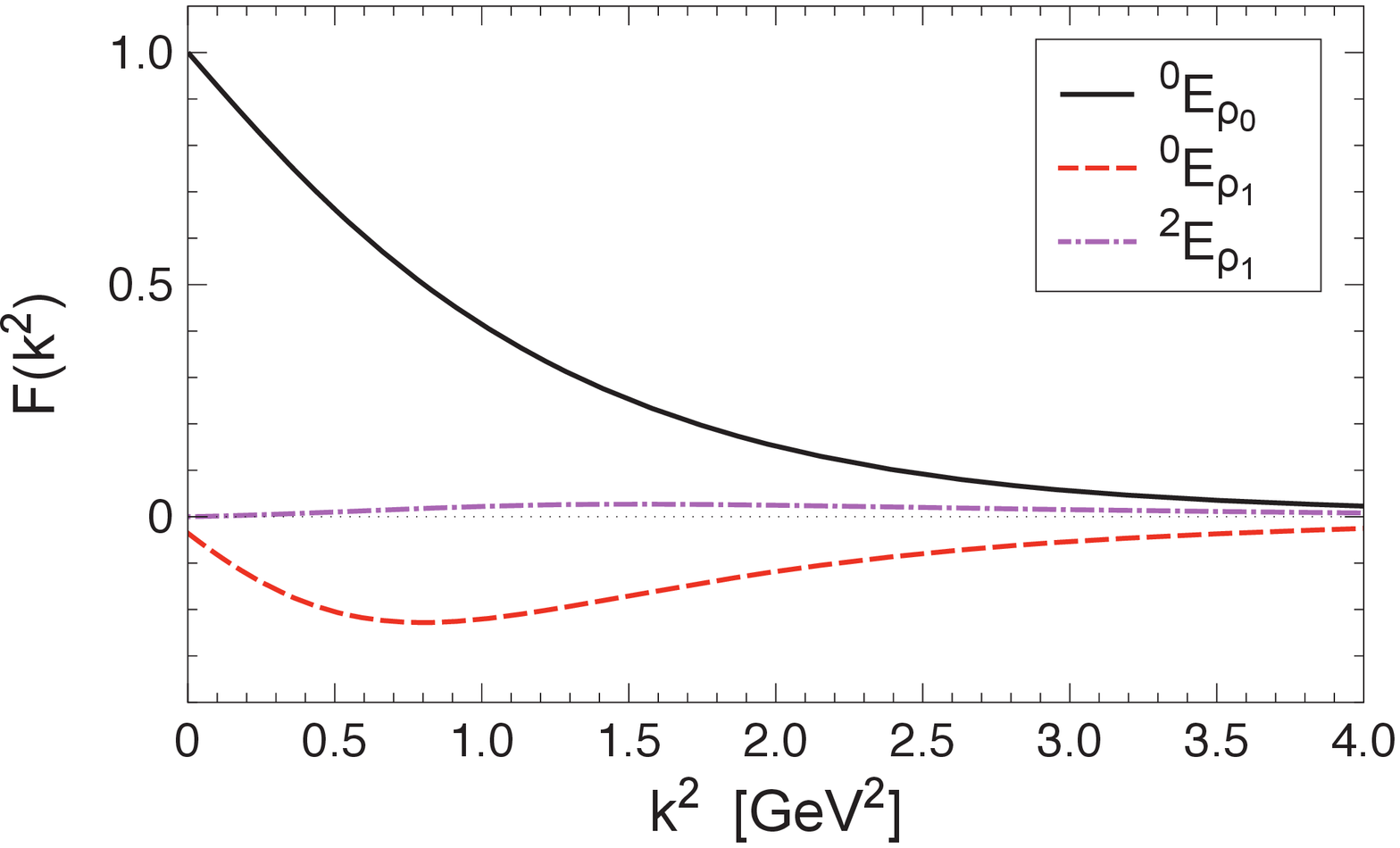}}
\end{minipage}\vspace*{3ex}

\begin{minipage}[t]{0.48\textwidth}
\leftline{\includegraphics[width=1.01\textwidth]{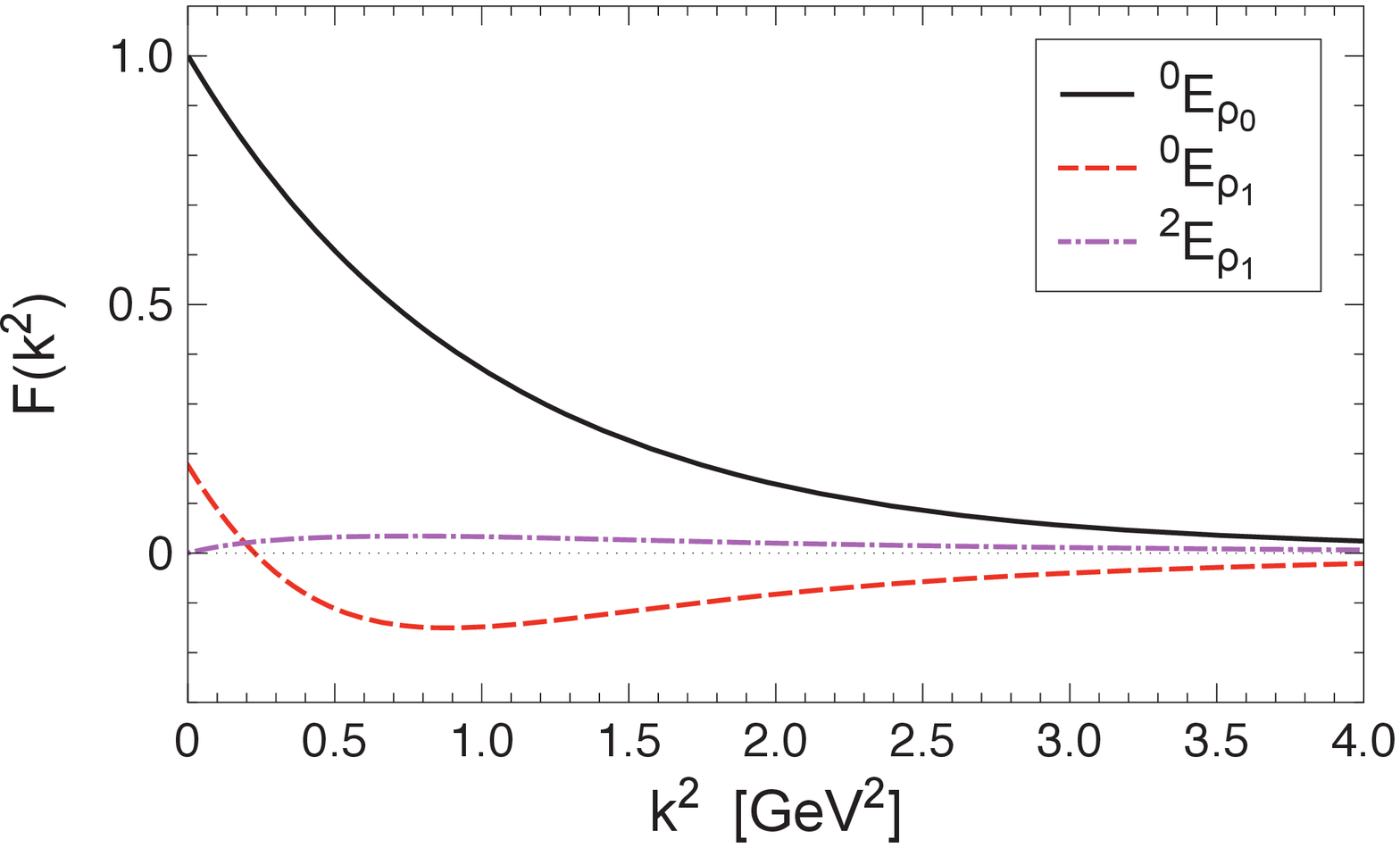}}
\end{minipage}
\begin{minipage}[t]{0.48\textwidth}
\rightline{\includegraphics[width=1.01\textwidth]{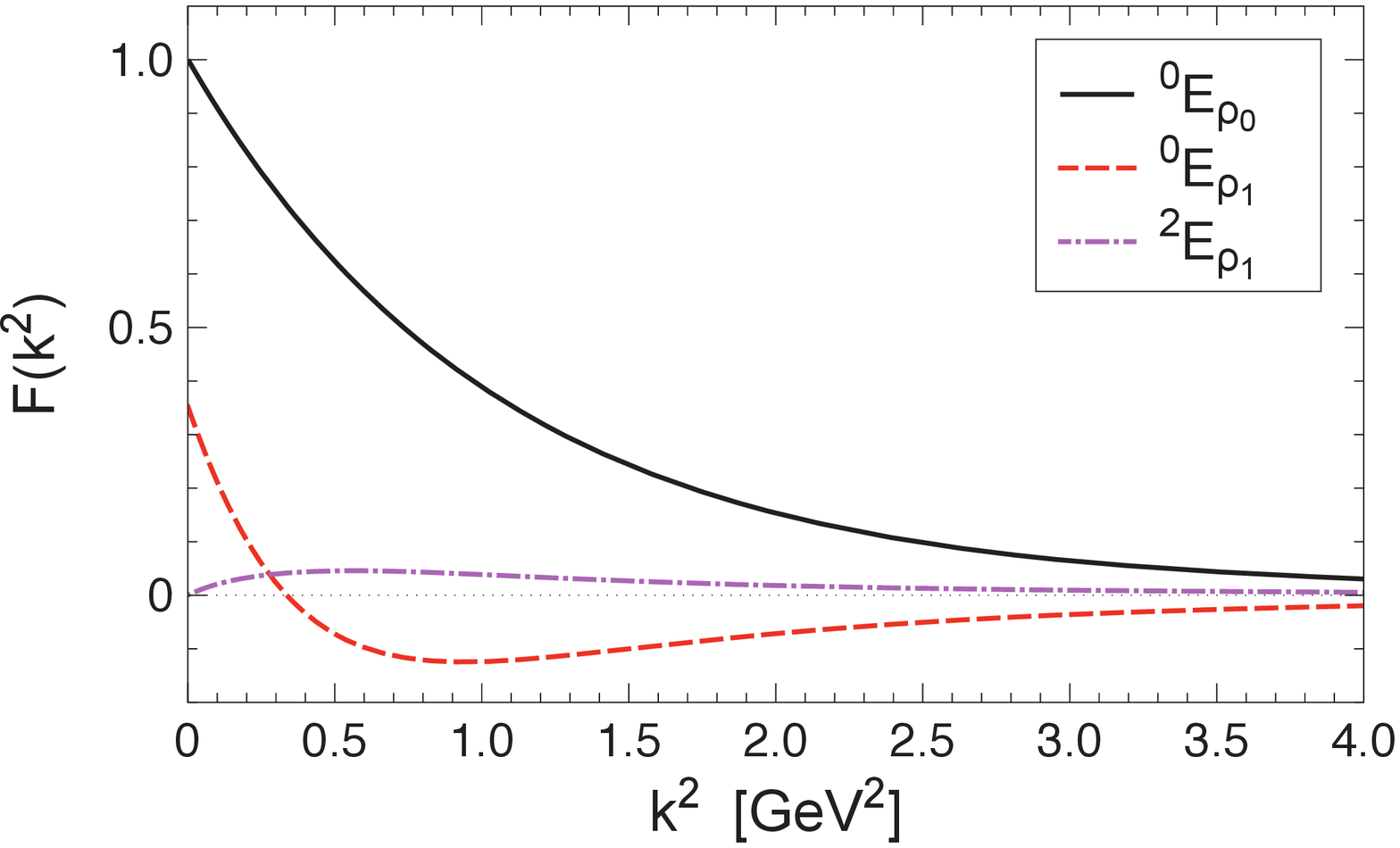}}
\end{minipage}\vspace*{3ex}

\begin{minipage}[t]{0.48\textwidth}
\leftline{\includegraphics[width=1.01\textwidth]{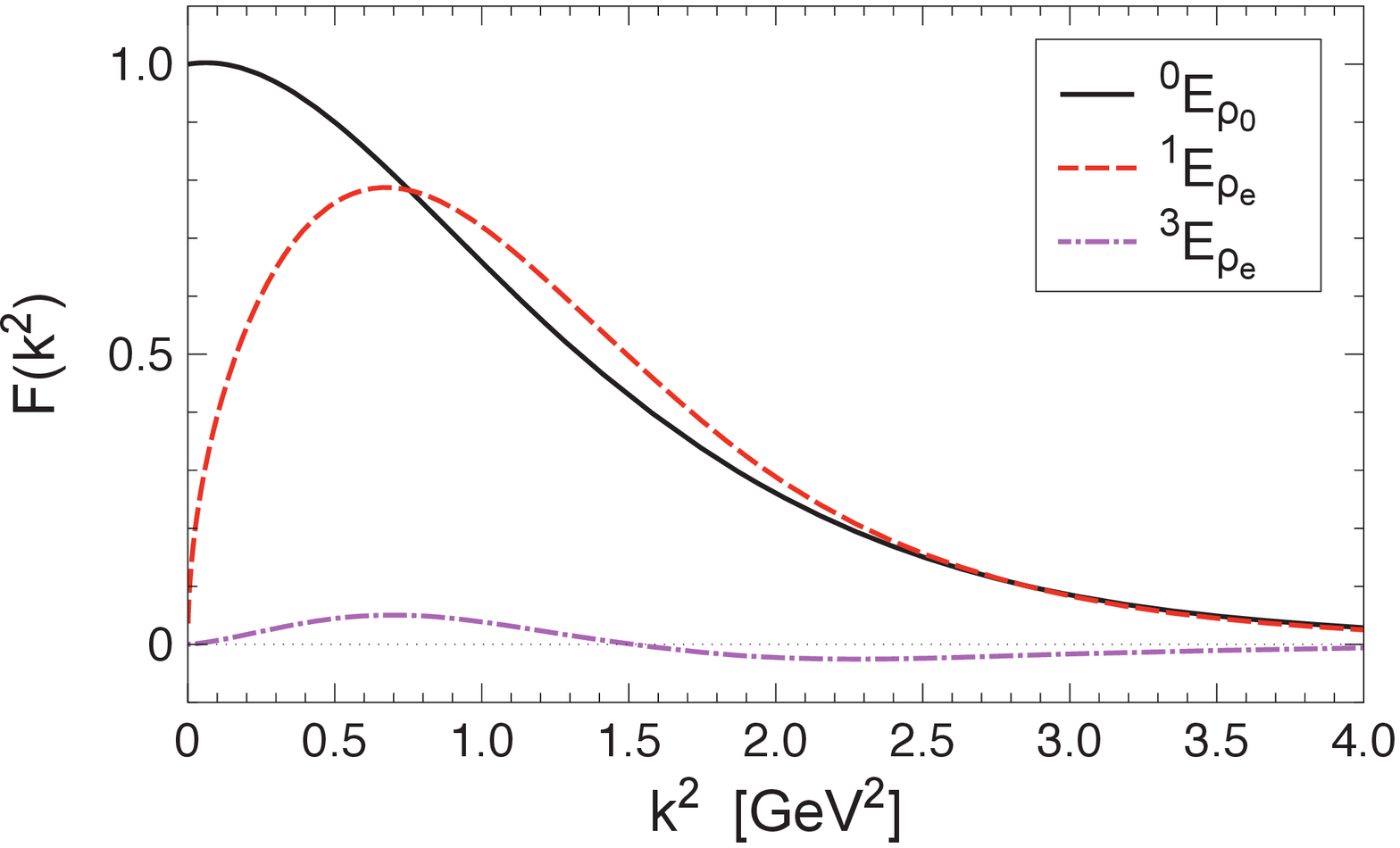}}
\end{minipage}
\begin{minipage}[t]{0.48\textwidth}
\rightline{\includegraphics[width=1.01\textwidth]{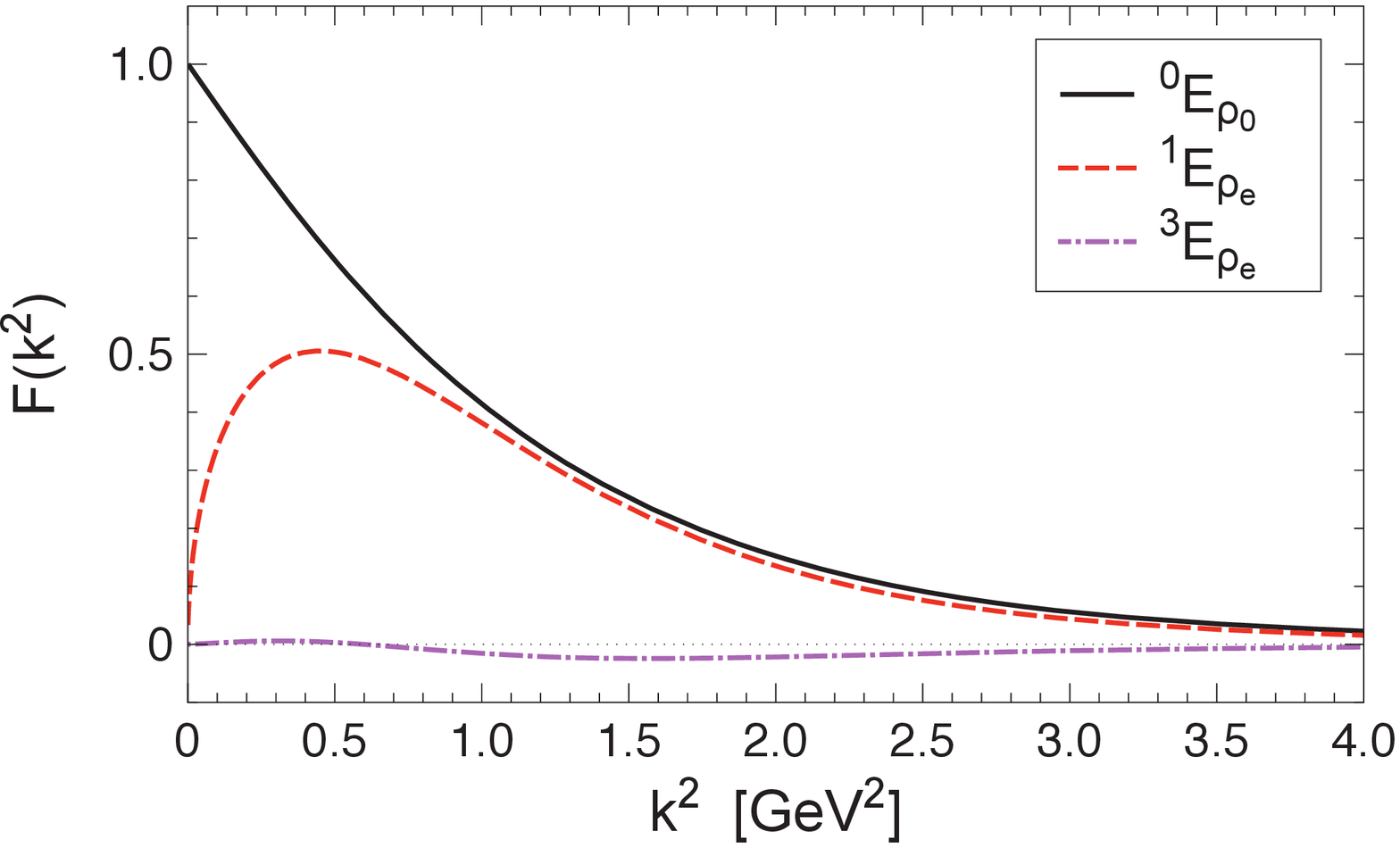}}
\end{minipage}\vspace*{3ex}

\begin{minipage}[t]{0.48\textwidth}
\leftline{\includegraphics[width=1.01\textwidth]{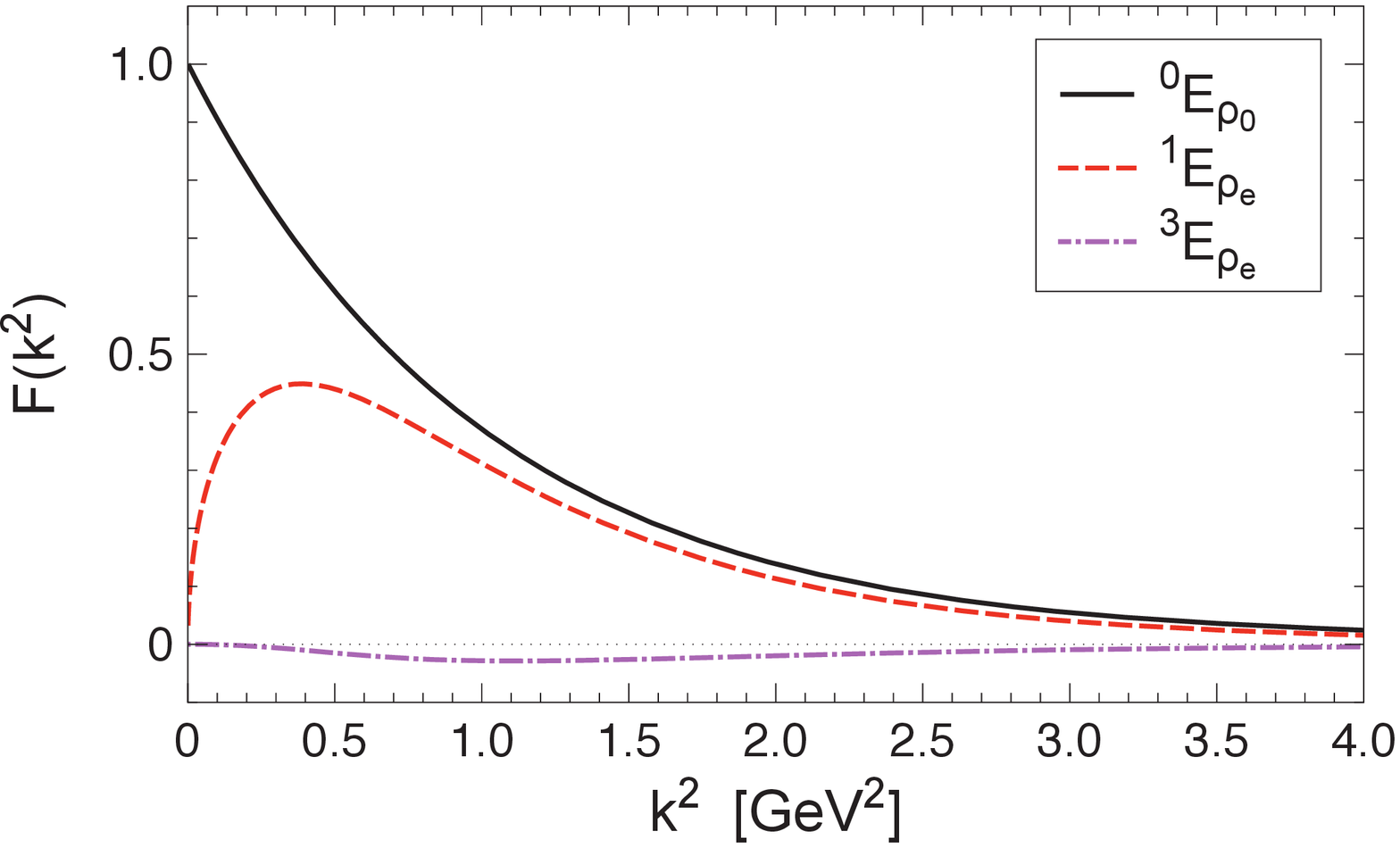}}
\end{minipage}
\begin{minipage}[t]{0.48\textwidth}
\rightline{\includegraphics[width=1.01\textwidth]{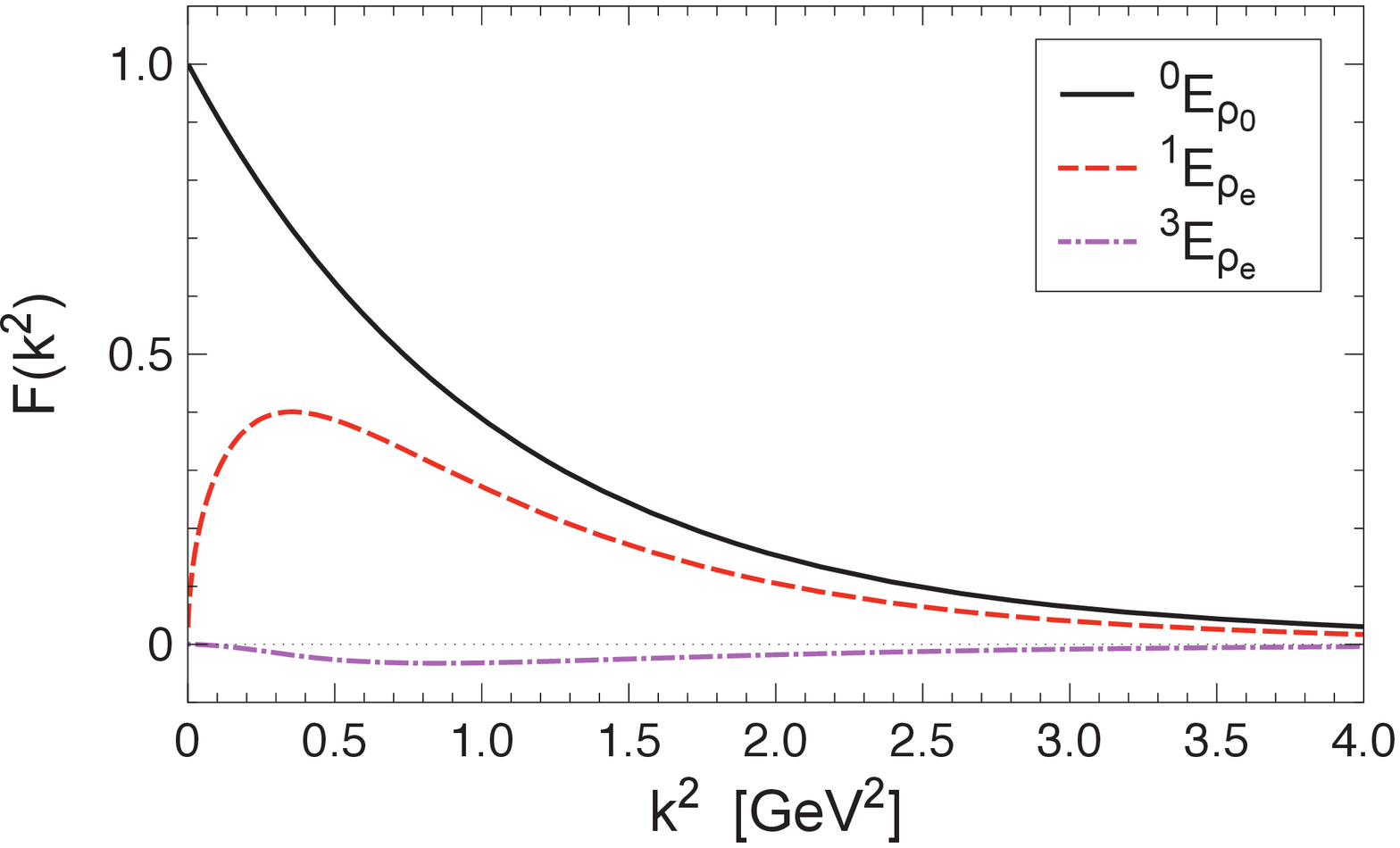}}
\end{minipage}
\end{minipage}

\caption{\label{rhoBSA} Vector mesons.  $\omega$-dependence of low-order Chebyshev-projections of leading invariant amplitude for ground-, radially-excited- and exotic-states: \emph{upper four panels}, ground and radial; \emph{lower four panels}, ground and exotic.  In all panels, \emph{solid} -- zeroth moment, ground-state; \emph{dashed} -- leading moment, comparison state; \emph{dash-dot} -- subleading moment, comparison state.  \emph{Row-1, left}, $\omega=0.4\,$GeV; \emph{Row-1, right}, $\omega=0.5\,$GeV; \emph{Row-2, left}, $\omega=0.6\,$GeV; and \emph{Row 2, right}, $\omega=0.7\,$GeV.  This pattern is repeated in the next two rows.  The normalisation is chosen such that $^0\!E_{\rho_0}(p^2=0)=1$; and $D\omega = (1.1\,{\rm GeV})^3$.}
\end{figure}

Whilst not directly observable, the momentum-dependence of meson Bethe-Salpeter amplitudes is a crucial determinative factor in the computation of measurable quantities.  In Figs.\,\ref{pionBSA} and \ref{rhoBSA}, therefore, we depict the $\omega$-dependence of a few low-order Chebyshev moments of the leading invariant amplitude for the pseudoscalar and vector mesons:
\begin{equation}
^n \!F_M(p^2) := \frac{2}{\pi} \int_{-1}^{1}\!\! dx\,\sqrt{1-x^2}\,U_n(x)\,F_M(k^2,x;P^2)\,,
\end{equation}
where $k\cdot P = x \sqrt{k^2 P^2}$ and $U_n(x)$ is a Chebyshev polynomial of the second kind.  NB.\ For pseudoscalar and vector states with natural $C$-parity, only the even moments are nonzero, whereas it is the odd moments which are nonzero for the exotic partners of these states.

The upper four panels in Fig.\,\ref{pionBSA} compare the amplitudes of the ground-state and first-radially-excited pseudoscalar mesons.  The ground-state is clearly insensitive to $\omega$.  However, as hoped for and anticipated, the radial excitation reacts strongly to variations in $\omega$.  Most notable is the suppression of $^0\!E_{\pi_1}$ with decreasing $\omega$, to be replaced by an increasingly large $^2\!E_{\pi_1}$.  Indeed, at $\omega=0.4\,$GeV, $^0\! E_{\pi_1}$ is almost negligible and possesses two zeros, instead of the single zero expected in the amplitude of a first radial excitation since the work of Ref.\,\cite{Holl:2004fr}.  In such circumstances, the radial excitation may even possess a smaller charge radius than the ground state \cite{Holl:2005vu}.

In our view these features signal that values of $\omega\lesssim 0.5\,$GeV in Eq.\,(\ref{CalGQC}) are unphysical; i.e., the long-range behaviour of a realistic $\beta$-function cannot dramatically suppress the radial excitation's leading amplitude nor induce it to have a second zero.  This perspective is supported by the following considerations.
Neither the homogeneous BSE nor the canonical normalisation condition fix the sign of  the Bethe-Salpeter amplitude at $k^2=0$.  As in quantum mechanics, this is arbitrary and cannot affect observables.  Another parallel with quantum mechanics is also relevant.  Namely, for a ground-state, the sign of the radial wave function at the origin in configuration space is the same as that of its analogue at the origin in momentum space, whereas these signs are opposite for the first radial excitation.  This pattern repeats for higher even- and odd-numbered radial excitations.
Here, a direct solution of the inhomogeneous BSE is instructive because this equation does determine signs.  For example, consider the pseudoscalar vertex: Fig.\,6 of Ref.\,\cite{Bhagwat:2007rj} illustrates a case in which the residue associated with the pseudoscalar meson ground-state is positive and that connected with the first radial excitation is negative, which is the behaviour found herein for $\omega \gtrsim 0.5\,$GeV.  The residue is a product of the pseudoscalar-meson's bound-state Bethe-Salpeter amplitude at $k^2=0$, $\Gamma_{0^-}(0;P^2)$, and $\rho_{0^-}$.  The latter is the expression in quantum field theory for the value of the Bethe-Salpeter wave function at the origin in configuration space.  Thus, the pattern exposed by the inhomogeneous BSE parallels that in quantum mechanics.

It is straightforward to see that this pattern is realised in the second, third and fourth panels of Fig.\,\ref{pionBSA}, which depict results obtained with $\omega\geq 0.5\,$GeV.  Therein, the $k^2=0$ values of the leading amplitudes' lowest Chebyshev projections are positive; and whilst that for the ground-state remains positive, that for the first radial excitation changes sign, so that it is a negative-definite function for $k^2\gtrsim 1\,$GeV$^2$.  In performing a Fourier transform, large-$k^2$ maps onto small $x^2$ and hence this behaviour guarantees that the Bethe-Salpeter wave function for the first radial excitation is negative at the origin in configuration space.

\begin{figure}[t]

\centerline{\includegraphics[clip,width=0.45\textwidth]{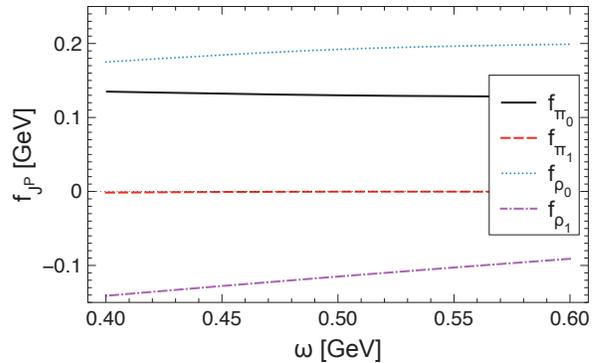}}

\caption{\label{Figleptonicdecay}
$\omega$-dependence of leptonic decay constants for pseudoscalar and vector mesons:
ground-state pion, solid line;
radially-excited pion, dashed line;
ground-state rho-meson, dotted line;
and radially-excited rho-meson, dash-dot line. ($D\omega = (1.1\,{\rm GeV})^3$.)}
\end{figure}

These observations reemphasise the peculiar character of the $\omega=0.4\,$GeV solution in the top-left panel of the Fig.\,\ref{pionBSA} and explain our choice of sign for all Bethe-Salpeter amplitudes.  The ground-state amplitude is positive at large-$k^2$, the first radial excitation is negative at large-$k^2$, and so on.  With this convention, one necessarily finds $\rho_{\pi_0}^\zeta>0$, $\rho_{\pi_1}^\zeta<0$, etc., and hence, from Eq.\,(\ref{gmorP}), $f_{\pi_0}>0$, $f_{\pi_1}<0$.  We depict the $\omega$-dependence of the leptonic decay constants in Fig.\,\ref{Figleptonicdecay}.

The bottom four panels of Fig.\,\ref{pionBSA} display low-order moments of the exotic-pseudoscalar-meson's leading invariant amplitude, contrasted with the ground-state's zeroth moment.  So long as $\omega \gtrsim 0.5\,$GeV, the first moment of the exotic amplitude is bounded above by $^0\! E_{\pi_0}$ and the third moment is negative definite.  This is the first time these features have been exposed but we expect them to be characteristic of the rainbow-ladder truncation.  It will be important to learn whether this pattern persists beyond rainbow-ladder truncation.

The top four panels in Fig.\,\ref{rhoBSA} compare the amplitudes of the ground-state and first-radially-excited vector mesons.  The ground-state is insensitive to $\omega$ so long as $\omega \gtrsim 0.5\,$GeV but again the radial excitation reacts strongly to variations in $\omega$.  In this case, natural behaviour for the excited state's amplitudes is only obtained for $\omega \gtrsim 0.6\,$GeV.  For smaller values, the zeroth moment is negative-definite and the second moment exhibits a zero.  NB.\ The sign of the amplitudes is fixed via the same prescription used for pseudoscalar mesons, and hence $f_{\rho_0}>0$, $f_{\rho_1}<0$.

The bottom four panels of Fig.\,\ref{rhoBSA} display low-order moments of the exotic-vector-meson's leading invariant amplitude, contrasted with the ground-state's zeroth moment.  In this case, so long as $\omega \gtrsim 0.6\,$GeV, the first moment of the exotic amplitude is bounded above by $^0\! E_{\rho_0}$ and the third moment is negative definite.
The similarity to the lower panels of Fig.\,\ref{pionBSA} encourages us in the expectation that these features are characteristic of the rainbow-ladder truncation.
Moreover, they suggest again that there is too much similarity between natural and exotic $C$-parity states in rainbow-ladder truncation.

In Fig.\,\ref{Figleptonicdecay} we depict the $\omega$-dependence of pseudoscalar- and vector-meson leptonic decay constants.  Those for the ground-states are positive whilst those for the first radial excitations are negative.  The origin of this outcome in an internally consistent treatment of bound-states was explained above.
Notable, too, is the small magnitude of the decay constant for the pion's first radial excitation: $f_{\pi_1}\approx -1\,$MeV.  This was predicted in Ref.\,\cite{Holl:2004fr} and is a consequence of the axial-vector Ward-Takahashi identity.  It is consistent with data on $\tau \to \pi(1300) \nu_\tau$  \cite{Diehl:2001xe} and numerical simulations of lattice-regularised QCD \cite{McNeile:2006qy}.

\section{Conclusion}
\label{epilogue}
Using an interaction kernel that is consonant with modern DSE- and lattice-QCD results, we employed a rainbow-ladder truncation of QCD's Dyson-Schwinger equations in an analysis of ground-state, radially-excited and exotic scalar-, vector- and flavoured-pseudoscalar-mesons.
We confirmed that rainbow-ladder truncation is incapable of providing realistic predictions for the masses of excited- and exotic-states; e.g., the ordering between pseudoscalar and vector radially-excited states is incorrect, and computed masses for exotic states are too low in comparison with other estimates.  Indeed, in rainbow-ladder truncation, it appears that exotic states are in most respects too much like their $C$-parity partners.

On the other hand, rainbow-ladder results do provide information that is useful in proceeding beyond this leading-order.
For example, in each channel the rainbow-ladder truncation indicates those invariant amplitudes which are likely to dominate in any solution of the Bethe-Salpeter equation.  This knowledge can be used in developing integral projection techniques that suppress ground-state contamination when searching for excited states.
Moreover, the response of observables, and the Bethe-Salpeter amplitudes which produce them, to changes in the infrared evolution of the interaction kernel can be used effectively to demarcate the domain of physically allowed possibilities for that evolution.  This is valuable in qualitatively constraining the long-range behaviour of QCD's $\beta$-function.
In addition, the symmetry-preserving character of the rainbow-ladder truncation and the ready access it provides to Bethe-Salpeter amplitudes for bound-states enable one to highlight and illustrate features of hadron observables that do not depend on details of the dynamics.

There are many indications that dynamical chiral symmetry breaking (DCSB), of which the momentum-dependence of the dressed-quark mass-function is a striking signal, has an enormous impact on hadron properties.  This study is one of a growing body which indicates that the veracious expression of DCSB in the bound-state problem is essential if one is to reliably predict and understand the spectrum and properties of excited and exotic hadrons.  Achieving this will provide the power to use extant and forthcoming data as a tool with which to chart the nonperturbative evolution of QCD's $\beta$-function.

\section*{Acknowledgments}
This work was supported by:
National Natural Science Foundation of China, under contract nos.~10705002 and 10935001;
and U.\,S.\ Department of Energy, Office of Nuclear Physics, contract no.~DE-AC02-06CH11357.


\begin{thebibliography}{99}

\bibitem{Wilson:1974sk}
  K.~G.~Wilson,
  Phys.\ Rev.\  D {\bf 10}, 2445 (1974).

\bibitem{Eichten:1978tg}
  E.~Eichten, K.~Gottfried, T.~Kinoshita, K.~D.~Lane and T.~M.~Yan,
  Phys.\ Rev.\  D {\bf 17}, 3090 (1978)
  [Erratum-ibid.\  D {\bf 21}, 313 (1980)].

\bibitem{Brambilla:2010cs}
  N.~Brambilla {\it et al.},
  Eur.\ Phys.\ J.\  C {\bf 71}, 1534 (2011).

\bibitem{Bali:2005fu}
  G.~S.~Bali, H.~Neff, T.~Duessel, T.~Lippert and K.~Schilling
  [SESAM Collaboration],
  {Phys.\ Rev.\  D} \textbf{71}, 114513 (2005). 

\bibitem{Chang:2009ae}
  L.~Chang, I.~C.~Clo\"et, B.~El-Bennich, T.~Kl\"ahn and C.~D.~Roberts,
  {Chin.\ Phys.\  C} \textbf{33}, pp.~1189-1196 (2009).

\bibitem{Krein:1990sf}
  G.~Krein, C.~D.~Roberts and A.~G.~Williams,
  {Int.\ J.\ Mod.\ Phys.\  A} \textbf{7}, 5607 (1992).

\bibitem{Roberts:1994dr}
  C.~D.~Roberts and A.~G.~Williams,
  Prog.\ Part.\ Nucl.\ Phys.  \textbf{33}, 477 (1994).

\bibitem{Roberts:2000aa}
  C.~D.~Roberts and S.~M.~Schmidt,
  Prog.\ Part.\ Nucl.\ Phys.\  {\bf 45}, S1 (2000).

\bibitem{Maris:2003vk}
  P.~Maris and C.~D.~Roberts,
  Int.\ J.\ Mod.\ Phys.\  E {\bf 12}, 297 (2003).

\bibitem{Pennington:2005be}
  M.~R.~Pennington,
  J.\ Phys.\ Conf.\ Ser.\  {\bf 18}, 1 (2005).

\bibitem{Roberts:2007ji}
  C.~D.~Roberts,
  Prog.\ Part.\ Nucl.\ Phys.  \textbf{61}, 50 (2008).

\bibitem{Chang:2011vu}
  L.~Chang, C.~D.~Roberts and P.~C.~Tandy,
  ``Selected highlights from the study of mesons,''
  arXiv:1107.4003 [nucl-th].

\bibitem{Roberts:2011rr}
  C.~D.~Roberts,
  ``Opportunities and Challenges for Theory in the N* program,''
  arXiv:1108.1030 [nucl-th].

\bibitem{Rodriguez-Quintero:2011}
    Ph.~Boucaud, J.~P.~Leroy, A.~Le~Yaouanc, J.~Micheli, O.~P\`{e}ne, J.~Rodr\'{i}guez-Quintero,
``The Infrared Behaviour of the Pure Yang-Mills Green Functions,''
arXiv:1109.1936 [hep-ph].

\bibitem{Bhagwat:2003vw}
  M.~S.~Bhagwat, M.~A.~Pichowsky, C.~D.~Roberts and P.~C.~Tandy,
  Phys.\ Rev.\  C \textbf{68}, 015203 (2003).

\bibitem{Bowman:2005vx}
  P.~O.~Bowman, U.~M.~Heller, D.~B.~Leinweber, M.~B.~Parappilly, A.~G.~Williams and J.~b.~Zhang,
  Phys.\ Rev.\  D \textbf{71}, 054507 (2005).

\bibitem{Bhagwat:2006tu} M.~S.~Bhagwat and P.~C.~Tandy,
  AIP Conf.\ Proc.\ \textbf{842}, 225 (2006).

\bibitem{Bhagwat:2007vx}
  M.~S.~Bhagwat, I.~C.~Clo\"et and C.~D.~Roberts,
  ``Covariance, Dynamics and Symmetries, and Hadron Form Factors,'' in \emph{Exclusive Reactions at High Momentum Transfer}, edited by A.~Radyushkin and P.~Stoler, World Scientific, Singapore, 2008, pp.~112-120; arXiv:0710.2059 [nucl-th].

\bibitem{Maris:1997tm}
  P.~Maris and C.~D.~Roberts,
  {Phys.\ Rev.\  C} \textbf{56}, 3369 (1997).

\bibitem{Holl:2004fr}
  A.~H\"oll, A.~Krassnigg and C.~D.~Roberts,
  {Phys.\ Rev.\  C} \textbf{70}, 042203(R) (2004).

\bibitem{Holl:2005vu}
  A.~H\"oll, A.~Krassnigg, P.~Maris, C.~D.~Roberts and S.~V.~Wright,
  {Phys.\ Rev.\  C} \textbf{71}, 065204 (2005).

\bibitem{Lucha:2006rq}
  W.~Lucha, D.~Melikhov and S.~Simula,
  Phys.\ Rev.\  D {\bf 74}, 054004 (2006).

\bibitem{McNeile:2006qy}
  C.~McNeile and C.~Michael  [UKQCD Collaboration],
  Phys.\ Lett.\  B {\bf 642}, 244 (2006).

\bibitem{Bhagwat:2007rj}
  M.~S.~Bhagwat, A.~H\"oll, A.~Krassnigg, C.~D.~Roberts and S.~V.~Wright,
  Few Body Syst.\  {\bf 40}, 209 (2007).

\bibitem{Chang:2009zb}
  L.~Chang and C.~D.~Roberts,
  Phys.\ Rev.\ Lett.\  {\bf 103}, 081601 (2009).

\bibitem{Fischer:2009jm}
  C.~S.~Fischer and R.~Williams,
  Phys.\ Rev.\ Lett.\  {\bf 103}, 122001 (2009).

\bibitem{Krassnigg:2009zh}
  A.~Krassnigg,
  Phys.\ Rev.\  D {\bf 80}, 114010 (2009).

\bibitem{Chang:2011ei}
  L.~Chang and C.~D.~Roberts,
  ``Tracing masses of ground-state light-quark mesons,''
  arXiv:1104.4821 [nucl-th].

\bibitem{Blank:2011qk}
  M.~Blank,
  ``Properties of quarks and mesons in the Dyson-Schwinger/Bethe-Salpeter approach,''
  arXiv:1106.4843 [hep-ph].

\bibitem{Qin:2011dd}
  S.~x.~Qin, L.~Chang, Y.~x.~Liu, C.~D.~Roberts and D.~J.~Wilson,
  ``Interaction model for the gap equation,''
  arXiv:1108.0603 [nucl-th].

\bibitem{Munczek:1994zz}
  H.~J.~Munczek,
  {Phys.\ Rev.\  D} \textbf{52}, 4736 (1995).

\bibitem{Bender:1996bb}
  A.~Bender, C.~D.~Roberts and L.~Von Smekal,
  {Phys.\ Lett.\  B} \textbf{380}, 7 (1996).

\bibitem{Chang:2010hb}
  L.~Chang, Y.~X.~Liu and C.~D.~Roberts,
  Phys.\ Rev.\ Lett.\  {\bf 106}, 072001 (2011).

\bibitem{Bloch:2002eq}
  J.~C.~R.~Bloch,
  Phys.\ Rev.\  D {\bf 66}, 034032 (2002).

\bibitem{Bashir:2009fv}
  A.~Bashir, A.~Raya, S.~Sanchez-Madrigal and C.~D.~Roberts,
  Few Body Syst.\  {\bf 46}, 229  (2009). 

\bibitem{Bashir:2011vg}
  A.~Bashir, A.~Raya and S.~Sanchez-Madrigal,
  ``Chiral Symmetry Breaking and Confinement Beyond Rainbow-Ladder
  Truncation,''
  arXiv:1108.4748 [hep-ph].

\bibitem{Cucchieri:2011aa}
  A.~Cucchieri, T.~Mendes, G.~M.~Nakamura and E.~M.~S.~Santos,
  AIP Conf.\ Proc.\  {\bf 1354}, 45  (2011). 

\bibitem{Flambaum:2005kc}
  V.~V.~Flambaum, A.~H\"oll, P.~Jaikumar, C.~D.~Roberts and S.~V.~Wright,
  Few Body Syst.\  {\bf 38}, 31 (2006).

\bibitem{Holl:2005st}
  A.~H\"oll, P.~Maris, C.~D.~Roberts and S.~V.~Wright,
  Nucl.\ Phys.\ Proc.\ Suppl.\  {\bf 161}, 87 (2006).

\bibitem{Bhagwat:2007ha}
  M.~S.~Bhagwat \emph{et~al}., 
  {Phys.\ Rev.\  C} \textbf{76}, 045203 (2007). 

\bibitem{LlewellynSmith:1969az}
  C.~H.~Llewellyn-Smith,
  Annals Phys.\  {\bf 53}, 521 (1969).

\bibitem{Bowman:2004jm}
  P.~O.~Bowman, U.~M.~Heller, D.~B.~Leinweber, M.~B.~Parappilly and A.~G.~Williams,
  Phys.\ Rev.\  D {\bf 70}, 034509 (2004). 

\bibitem{Aguilar:2009nf}
  A.~C.~Aguilar, D.~Binosi, J.~Papavassiliou and J.~Rodriguez-Quintero,
  Phys.\ Rev.\  D {\bf 80}, 085018 (2009). 

\bibitem{Aguilar:2010gm}
  A.~C.~Aguilar, D.~Binosi and J.~Papavassiliou,
  JHEP {\bf 1007}, 002 (2010).

\bibitem{Skullerud:2003qu}
  J.~I.~Skullerud, P.~O.~Bowman, A.~Kizilersu, D.~B.~Leinweber and A.~G.~Williams,
  JHEP {\bf 0304}, 047 (2003). 

\bibitem{Bhagwat:2004kj}
  M.~S.~Bhagwat and P.~C.~Tandy,
  Phys.\ Rev.\  D {\bf 70}, 094039 (2004). 

\bibitem{Krassnigg:2009gd}
  A.~Krassnigg,
  PoS \textbf{CONFINEMENT\,8}, 075 (2008). 

\bibitem{Cloet:2008fw}
  I.~C.~Cloet and C.~D.~Roberts,
  PoS {\bf LC2008}, 047 (2008).

\bibitem{Nakamura:2010zzi}
  K.~Nakamura {\it et al.}, 
  J.\ Phys.\ G {\bf 37} 075021 (2010). 

\bibitem{Roberts:2011IHC}
    L.~Chang, C.\,D.~Roberts and P.\,C.~Tandy,
    ``Expanding the concept of in-hadron condensates,'' arXiv:1109.2903 [nucl-th].

\bibitem{Ivanov:1998ms}
  M.~A.~Ivanov, Yu.~L.~Kalinovsky and C.~D.~Roberts,
  Phys.\ Rev.\  D {\bf 60}, 034018 (1999).

\bibitem{Pelaez:2006nj}
  J.~R.~Pelaez and G.~Rios,
  Phys.\ Rev.\ Lett.\  {\bf 97}, 242002 (2006). 

\bibitem{RuizdeElvira:2010cs}
  J.~Ruiz de Elvira, J.~R.~Pelaez, M.~R.~Pennington and D.~J.~Wilson,
  ``Chiral Perturbation Theory, the ${1/N_c}$ expansion and Regge behaviour determine the structure of the lightest scalar meson,''
  arXiv:1009.6204 [hep-ph].

\bibitem{Brodsky:2010xf}
  S.~J.~Brodsky, C.~D.~Roberts, R.~Shrock and P.~C.~Tandy,
  Phys.\ Rev.\  C {\bf 82}, 022201(R) (2010).

\bibitem{Burden:2002ps}
  C.~J.~Burden and M.~A.~Pichowsky,
  Few Body Syst.\  {\bf 32}, 119 (2002). 

\bibitem{Eichmann:2008ae}
  G.~Eichmann, R.~Alkofer, I.~C.~Clo\"et, A.~Krassnigg and C.~D.~Roberts,
  {Phys.\ Rev.\  C} \textbf{77}, 042202(R) (2008).

\bibitem{Roberts:2011cf}
  H.~L.~L.~Roberts, L.~Chang, I.~C.~Clo\"et and C.~D.~Roberts,
  Few Body Syst.\  {\bf 51}, 1 (2011). 

\bibitem{Maris:1999nt}
  P.~Maris and P.~C.~Tandy,
  Phys.\ Rev.\  C {\bf 60}, 055214 (1999).

\bibitem{Roberts:2011wy}
  H.~L.~L.~Roberts, A.~Bashir, L.~X.~Guti\'{e}rrez-Guerrero, C.~D.~Roberts and D.~J.~Wilson,
  Phys.\ Rev.\ C \textbf{83}, 065206 (2011). 

\bibitem{Dudek:2011bn}
  J.~J.~Dudek,
  ``The lightest hybrid meson supermultiplet in QCD,''
  arXiv:1106.5515 [hep-ph].

\bibitem{Diehl:2001xe}
  M.~Diehl and G.~Hiller,
  JHEP {\bf 0106}, 067 (2001).

\end{thebibliography}
\end{document}